\documentclass[screen=true, sigplan]{acmart}
\usepackage[utf8]{inputenc}
\usepackage[T1]{fontenc}

\newcommand{\packageGraphicx}{\usepackage{graphicx}}
\newcommand{\packageHyperref}{\usepackage{hyperref}}
\newcommand{\renewrmdefault}{\renewcommand{\rmdefault}{ptm}}
\newcommand{\packageRelsize}{\usepackage{relsize}}
\newcommand{\packageAmsmath}{\usepackage{amsmath}}
\newcommand{\packageMathabx}{\usepackage{mathabx}}
\newcommand{\packageWasysym}{
  \let\leftmoon\relax \let\rightmoon\relax \let\fullmoon\relax \let\newmoon\relax \let\diameter\relax
  \usepackage[nointegrals]{wasysym}}
\newcommand{\packageTxfonts}{
  \let\widering\relax
  \let\oldwidebar\widebar
  \let\widebar\relax
  \usepackage{newtxmath}
  \ifx\widebar\relax
    \let\widebar\oldwidebar
  \fi
}
\newcommand{\packageTextcomp}{\usepackage{textcomp}}
\newcommand{\packageFramed}{\usepackage{framed}}
\newcommand{\packageHyphenat}{\usepackage[htt]{hyphenat}}
\newcommand{\packageColor}{\usepackage[usenames,dvipsnames]{color}}
\newcommand{\doHypersetup}{\hypersetup{bookmarks=true,bookmarksopen=true,bookmarksnumbered=true}}
\newcommand{\packageTocstyle}{}
\newcommand{\packageCJK}{\IfFileExists{CJK.sty}{\usepackage{CJK}}{}}
\renewcommand\packageColor\relax
\renewcommand\packageTocstyle\relax
\renewcommand\packageMathabx{\ifx\bigtimes\undefined \usepackage{mathabx} \else \relax \fi}
\renewcommand\packageTxfonts\relax

\renewcommand{\renewrmdefault}{}

\packageGraphicx
\packageHyperref
\renewrmdefault
\packageRelsize
\packageAmsmath
\packageMathabx
\packageWasysym
\packageTxfonts
\packageTextcomp
\packageFramed
\packageHyphenat
\packageColor
\doHypersetup
\packageTocstyle
\packageCJK

\newcommand{\sectionNewpage}{}

\newcommand{\preDoc}{}
\newcommand{\postDoc}{}

\newcommand{\BookRef}[2]{\emph{#2}}
\newcommand{\ChapRef}[2]{\SecRef{#1}{#2}}
\newcommand{\SecRef}[2]{section~#1}
\newcommand{\PartRef}[2]{part~#1}
\newcommand{\BookRefUC}[2]{\BookRef{#1}{#2}}
\newcommand{\ChapRefUC}[2]{\SecRefUC{#1}{#2}}
\newcommand{\SecRefUC}[2]{Section~#1}
\newcommand{\PartRefUC}[2]{Part~#1}

\newcommand{\BookRefLocal}[3]{\hyperref[#1]{\BookRef{#2}{#3}}}
\newcommand{\ChapRefLocal}[3]{\hyperref[#1]{\ChapRef{#2}{#3}}}
\newcommand{\SecRefLocal}[3]{\hyperref[#1]{\SecRef{#2}{#3}}}
\newcommand{\PartRefLocal}[3]{\hyperref[#1]{\PartRef{#2}{#3}}}
\newcommand{\BookRefLocalUC}[3]{\hyperref[#1]{\BookRefUC{#2}{#3}}}
\newcommand{\ChapRefLocalUC}[3]{\hyperref[#1]{\ChapRefUC{#2}{#3}}}
\newcommand{\SecRefLocalUC}[3]{\hyperref[#1]{\SecRefUC{#2}{#3}}}
\newcommand{\PartRefLocalUC}[3]{\hyperref[#1]{\PartRefUC{#2}{#3}}}

\newcommand{\BookRefUN}[1]{\BookRef{}{#1}}

\newcommand{\SecRefUN}[1]{``#1''}

\newcommand{\BookRefLocalUN}[2]{\hyperref[#1]{\BookRefUN{#2}}}

\newcommand{\SecRefLocalUN}[2]{\hyperref[#1]{\SecRefUN{#2}}}

\newcommand{\SectionNumberLink}[2]{\hyperref[#1]{#2}}

\newcommand{\Scribtexttt}[1]{{\texttt{#1}}}

\newcommand{\planetName}[1]{PLane\hspace{-0.1ex}T}

\newcommand{\Stttextmore}{{\fontencoding{T1}\selectfont>}}
\newcommand{\Stttextless}{{\fontencoding{T1}\selectfont<}}

\makeatletter
\def\empty@finalstrut#1{%
  \unskip\ifhmode\nobreak\fi\vrule\@width\z@\@height\z@\@depth\z@}
\def\no@strut{\global\setbox\@arstrutbox\hbox{%
    \vrule \@height\z@
           \@depth\z@
           \@width\z@}%
    \gdef\@endpbox{\empty@finalstrut\@arstrutbox\par\egroup\hfil}%
}%
\def\yes@strut{\global\setbox\@arstrutbox\hbox{%
    \vrule \@height\arraystretch \ht\strutbox
           \@depth\arraystretch \dp\strutbox
           \@width\z@}%
    \gdef\@endpbox{\@finalstrut\@arstrutbox\par\egroup\hfil}%
}%
\def\@mkpream#1{\@firstamptrue\@lastchclass6
  \let\@preamble\@empty\def\empty@preamble{\add@ins}%
  \let\protect\@unexpandable@protect
  \let\@sharp\relax\let\add@ins\relax
  \let\@startpbox\relax\let\@endpbox\relax
  \@expast{#1}%
  \expandafter\@tfor \expandafter
    \@nextchar \expandafter:\expandafter=\reserved@a\do
       {\@testpach\@nextchar
    \ifcase \@chclass \@classz \or \@classi \or \@classii \or \@classiii
      \or \@classiv \or\@classv \fi\@lastchclass\@chclass}%
  \ifcase \@lastchclass \@acol
      \or \or \@preamerr \@ne\or \@preamerr \tw@\or \or \@acol \fi}
\def\@addamp{%
  \if@firstamp
    \@firstampfalse
    \edef\empty@preamble{\add@ins}%
  \else
    \edef\@preamble{\@preamble &}%
    \edef\empty@preamble{\expandafter\noexpand\empty@preamble &\add@ins}%
  \fi}
\newif\iftw@hlines \tw@hlinesfalse
\def\@xhline{\ifx\reserved@a\hline
               \tw@hlinestrue
             \else\ifx\reserved@a\Hline
               \tw@hlinestrue
             \else
               \tw@hlinesfalse
             \fi\fi
      \iftw@hlines
        \aftergroup\do@after
      \fi
      \ifnum0=`{\fi}%
}
\def\do@after{\emptyrow[\the\doublerulesep]}
\def\emptyrow{\noalign\bgroup\@ifnextchar[\@emptyrow{\@emptyrow[\z@]}}
\def\@emptyrow[#1]{\no@strut\gdef\add@ins{\vrule \@height\z@ \@depth#1 \@width\z@}\egroup%
\empty@preamble\\
\noalign{\yes@strut\gdef\add@ins{\vrule \@height\z@ \@depth\z@ \@width\z@}}%
}
\def\tabrow#1{\noalign\bgroup\@ifnextchar[{\@tabrow{#1}}{\@tabrow{#1}[]}}
\def\@tabrow#1[#2]{\no@strut\egroup#1\ifx.#2.\\\else\\[#2]\fi\noalign{\yes@strut}}
\def\endpltstabular{\crcr\egroup\egroup \egroup}
\expandafter \let \csname endpltstabular*\endcsname = \endpltstabular
\def\pltstabular{\let\@halignto\@empty\@pltstabular}
\@namedef{pltstabular*}#1{\def\@halignto{to#1}\@pltstabular}
\def\@pltstabular{\leavevmode \bgroup \let\@acol\@tabacol
   \let\@classz\@tabclassz
   \let\@classiv\@tabclassiv \let\\\@tabularcr\@stabarray}
\def\@stabarray{\m@th\@ifnextchar[\@sarray{\@sarray[c]}}
\def\@sarray[#1]#2{%
  \bgroup
  \setbox\@arstrutbox\hbox{%
    \vrule \@height\arraystretch\ht\strutbox
           \@depth\arraystretch \dp\strutbox
           \@width\z@}%
  \@mkpream{#2}%
  \edef\@preamble{%
    \ialign \noexpand\@halignto
      \bgroup \@arstrut \@preamble \tabskip\z@skip \cr}%
  \let\@startpbox\@@startpbox \let\@endpbox\@@endpbox
  \let\tabularnewline\\%
    \let\@sharp##%
    \set@typeset@protect
    \lineskip\z@skip\baselineskip\z@skip
    \@preamble}

\makeatother

\newenvironment{bigtabular}{\begin{pltstabular}}{\end{pltstabular}}

\newlength{\stabLeft}
\newcommand{\bigtableleftpad}{\hspace{\stabLeft}}

\newenvironment{SingleColumn}{\begin{list}{}{\topsep=0pt\partopsep=0pt%
\listparindent=0pt\itemindent=0pt\labelwidth=0pt\leftmargin=0pt\rightmargin=0pt%
\itemsep=0pt\parsep=0pt}\item}{\end{list}}

\newcommand{\SCodePreSkip}{\vskip\abovedisplayskip}
\newcommand{\SCodePostSkip}{\vskip\belowdisplayskip}
\newenvironment{SCodeFlow}{\SCodePreSkip\begin{list}{}{\topsep=0pt\partopsep=0pt%
\listparindent=0pt\itemindent=0pt\labelwidth=0pt\leftmargin=2ex\rightmargin=2ex%
\itemsep=0pt\parsep=0pt}\item}{\end{list}\SCodePostSkip}

\newcommand{\SVInsetPreSkip}{\vskip\abovedisplayskip}
\newcommand{\SVInsetPostSkip}{\vskip\belowdisplayskip}

\newcommand{\titleAndVersionAndAuthors}[3]{\title{#1\\{\normalsize \SVersionBefore{}#2}}\author{#3}\maketitle}

\newcommand{\titleAndEmptyVersionAndAuthors}[3]{\title{#1}\author{#3}\maketitle}

\newcommand{\SAuthor}[1]{#1}
\newcommand{\SAuthorSep}[1]{\qquad}
\newcommand{\SVersionBefore}[1]{Version }

\newcommand{\SNumberOfAuthors}[1]{}

\let\SOriginalthesubsection\thesubsection
\let\SOriginalthesubsubsection\thesubsubsection

\newcommand{\Ssection}[2]{\section[#1]{#2}\let\thesubsection\SOriginalthesubsection}
\newcommand{\Ssubsection}[2]{\subsection[#1]{#2}\let\thesubsubsection\SOriginalthesubsubsection}

\newcommand{\Ssectionstar}[1]{\section*{#1}\renewcommand*\thesubsection{\arabic{subsection}}\setcounter{subsection}{0}}

\newcommand{\Ssectionstarx}[2]{\Ssectionstar{#2}\phantomsection\addcontentsline{toc}{section}{#1}}

\newcounter{GrouperTemp}

\newcommand{\Snolinkurl}[1]{\nolinkurl{#1}}

\newcommand{\SAuthorinfo}[4]{#1}
\newcommand{\SAuthorPlace}[1]{#1}
\newcommand{\SAuthorEmail}[1]{#1}

\newcommand{\SConferenceInfo}[2]{}
\newcommand{\SCopyrightYear}[1]{}
\newcommand{\SCopyrightData}[1]{}
\newcommand{\Sdoi}[1]{}

\newcommand{\SCategory}[3]{}
\newcommand{\SCategoryPlus}[4]{}
\newcommand{\STerms}[1]{}
\newcommand{\SKeywords}[1]{}

\newcommand{\AutobibLink}[1]{\color{ACMPurple}{#1}}
\setcopyright{acmlicensed}
\acmPrice{15.00}
\acmDOI{10.1145/3609025.3609478}
\acmYear{2023}
\copyrightyear{2023}
\acmSubmissionID{icfpws23funarchmain-id4-p}
\acmISBN{979-8-4007-0297-6/23/09}
\acmConference[FUNARCH '23]{Proceedings of the 1st ACM SIGPLAN International Workshop on Functional Software Architecture}{September 8, 2023}{Seattle, WA, USA}
\acmBooktitle{Proceedings of the 1st ACM SIGPLAN International Workshop on Functional Software Architecture (FUNARCH '23), September 8, 2023, Seattle, WA, USA}
\received{2023-06-01}
\received[accepted]{2023-06-28}

\newcommand{\Autobibtarget}[1]{\phantomsection#1}

\usepackage{calc}
\newlength{\ABcollength}
\newcommand{\Autocolbibnumber}[1]{\parbox[t]{5ex}{\hfill#1~~\vspace{1.0ex}}}
\newcommand{\Autocolbibentry}[1]{\setlength{\ABcollength}{\linewidth-5ex}\parbox[t]{\ABcollength}{#1\vspace{1.0ex}}}

\newcommand{\Autobibref}[1]{#1}

\providecommand{\AutobibLink}[1]{#1}

\newcommand{\pseudodoi}[1]{#1}
\usepackage{ccaption}

\makeatletter
\@ifundefined{belowcaptionskip}{}{}
\makeatother

\newcommand{\Legend}[1]{~

                        \hrule width \hsize height .33pt
                        \vspace{4pt}
                        \legend{#1}}

\newcommand{\FigureTarget}[2]{#1}
\newcommand{\FigureRef}[2]{#1}

\newlength{\FigOrigskip}
\FigOrigskip=\parskip

\newcommand{\FigureSetRef}{\refstepcounter{figure}}

\newenvironment{Figure}{\begin{figure}\FigureSetRef}{\end{figure}}
\newenvironment{FigureMulti}{\begin{figure*}[t!p]\FigureSetRef}{\end{figure*}}

\newenvironment{Centerfigure}{\begin{Xfigure}\centering\item}{\end{Xfigure}}

\newenvironment{Xfigure}{\begin{list}{}{\leftmargin=0pt\topsep=0pt\parsep=\FigOrigskip\partopsep=0pt}}{\end{list}}

\newenvironment{FigureInside}{}{}

\newcommand{\Centertext}[1]{\begin{center}#1\end{center}}

\newcommand{\SColorize}[2]{\color{#1}{#2}}

\newcommand{\SHyphen}[1]{#1}

\newcommand{\inColor}[2]{{\SHyphen{\Scribtexttt{\SColorize{#1}{#2}}}}}
\definecolor{PaleBlue}{rgb}{0.90,0.90,1.0}
\definecolor{LightGray}{rgb}{0.90,0.90,0.90}
\definecolor{CommentColor}{rgb}{0.76,0.45,0.12}
\definecolor{ParenColor}{rgb}{0.52,0.24,0.14}
\definecolor{IdentifierColor}{rgb}{0.15,0.15,0.50}
\definecolor{ResultColor}{rgb}{0.0,0.0,0.69}
\definecolor{ValueColor}{rgb}{0.13,0.55,0.13}
\definecolor{OutputColor}{rgb}{0.59,0.00,0.59}

\newcommand{\RktPn}[1]{\inColor{ParenColor}{#1}}

\newcommand{\RktSym}[1]{\inColor{IdentifierColor}{#1}}

\newcommand{\RktVal}[1]{\inColor{ValueColor}{#1}}

\newcommand{\RktModLink}[1]{\inColor{blue}{#1}}

\newcommand{\RktMod}[1]{\inColor{black}{#1}}

\newenvironment{RktBlk}{}{}

\newcommand{\RBackgroundLabel}[1]{}

\newcommand{\NoteBox}[1]{\footnote{#1}}
\newcommand{\NoteContent}[1]{#1}

\newcommand{\FootnoteRef}[1]{}
\newcommand{\FootnoteTarget}[1]{}

\newcommand{\FootnoteBlockContent}[1]{}

\renewcommand{\titleAndVersionAndAuthors}[3]{\title{#1}#3\maketitle}
\renewcommand{\titleAndEmptyVersionAndAuthors}[3]{\titleAndVersionAndAuthors{#1}{#2}{#3}}

\def\SAuthor#1{\SAutoAuthor#1\SAutoAuthorDone{#1}}
\def\SAutoAuthorDone#1{}
\def\SAutoAuthor{\futurelet\next\SAutoAuthorX}
\def\SAutoAuthorX{\ifx\next\SAuthorinfo \let\Snext\relax \else \let\Snext\SToAuthorDone \fi \Snext}
\def\SToAuthorDone{\futurelet\next\SToAuthorDoneX}
\def\SToAuthorDoneX#1{\ifx\next\SAutoAuthorDone \let\Snext\SAddAuthorInfo \else \let\Snext\SToAuthorDone \fi \Snext}
\newcommand{\SAddAuthorInfo}[1]{\SAuthorinfo{#1}{}{}}

\renewcommand{\SAuthorinfo}[4]{\author{#1}{#2}{#3}{#4}}
\renewcommand{\SAuthorSep}[1]{}

\renewcommand{\SAuthorPlace}[1]{\affiliation{#1}}
\renewcommand{\SAuthorEmail}[1]{\email{#1}}

\renewcommand{\SConferenceInfo}[2]{\conferenceinfo{#1}{#2}}
\renewcommand{\SCopyrightData}[1]{\copyrightdata{#1}}

\renewcommand{\SCategory}[3]{\category{#1}{#2}{#3}}
\renewcommand{\SCategoryPlus}[4]{\category{#1}{#2}{#3}[#4]}
\renewcommand{\STerms}[1]{\terms{#1}}
\renewcommand{\SKeywords}[1]{\keywords{#1}}

\newcommand{\SccsdescNumber}[2]{\ccsdesc[#1]{#2}}
\renewcommand{\Legend}[1]{\legend{#1}}
\begin{document}
\preDoc

\begin{abstract}Some object{-}oriented GUI toolkits tangle state management with
rendering. Functional shells and observable toolkits like GUI Easy
simplify and promote the creation of reusable views by analogy to
functional programming. We have successfully used GUI Easy on small and
large GUI projects. We report on our experience constructing and using
GUI Easy and derive from that experience several architectural patterns
and principles for building functional programs out of imperative
systems.\end{abstract}

\SccsdescNumber{500}{Software and its engineering~Publish-subscribe / event-based architectures}

\SccsdescNumber{500}{Software and its engineering~Reusability}

\SccsdescNumber{300}{Software and its engineering~Classes and objects}

\SccsdescNumber{300}{Software and its engineering~Extensible languages}

\titleAndEmptyVersionAndAuthors{Functional Shell and Reusable Components for Easy GUIs}{}{\SNumberOfAuthors{2}\SAuthor{\SAuthorinfo{D. Ben Knoble}{}{\SAuthorPlace{\institution{Independent}\city{Richmond}\state{Virginia}\country{USA}}}{\SAuthorEmail{ben.knoble+funarch2023@gmail.com}}}\SAuthorSep{}\SAuthor{\SAuthorinfo{Bogdan Popa}{}{\SAuthorPlace{\institution{Independent}\city{Cluj{-}Napoca}\state{Cluj}\country{Romania}}}{\SAuthorEmail{bogdan@defn.io}}}}
\label{t:x28part_x22Functionalx5fShellx5fandx5fReusablex5fComponentsx5fforx5fEasyx5fGUIsx22x29}

\noindent 

\noindent 

\noindent

\sectionNewpage

\Ssection{Introduction}{Introduction}\label{t:x28part_x22Introductionx22x29}

Object{-}oriented programming is traditionally considered a good paradigm
for building graphical (GUI) programs due to inheritance, composition,
and specialization. Racket{'}s GUI toolkit\Autobibref{~[\hyperref[t:x28autobib_x22Matthew_Flattx2c_Robert_Bruce_Findlerx2c_and_John_ClementsGUIx3a_Racket_Graphics_ToolkitPLT_Design_Incx2ex2c_PLTx2dTRx2d2010x2d32010httpsx3ax2fx2fracketx2dlangx2eorgx2ftr3x2fx22x29]{\AutobibLink{12}}]} is
object{-}oriented, with message{-}passing widgets and mutable state. The
Racket platform\Autobibref{~[\hyperref[t:x28autobib_x22Matthew_Flatt_and_PLTReferencex3a_RacketPLT_Design_Incx2ex2c_PLTx2dTRx2d2010x2d12010httpsx3ax2fx2fracketx2dlangx2eorgx2ftr1x2fx22x29]{\AutobibLink{14}}]} provides the core class and object
library for the GUI toolkit.

\begin{Figure}\begin{Centerfigure}\begin{FigureInside}\begin{RktBlk}\begin{SingleColumn}\RktModLink{\RktMod{\#lang}}\mbox{\hphantom{\Scribtexttt{x}}}\RktModLink{\RktSym{racket/gui}}

\RktPn{(}\RktSym{define}\mbox{\hphantom{\Scribtexttt{x}}}\RktSym{f}\mbox{\hphantom{\Scribtexttt{x}}}\RktPn{(}\RktSym{new}\mbox{\hphantom{\Scribtexttt{x}}}\RktSym{frame\%}\mbox{\hphantom{\Scribtexttt{x}}}\RktPn{[}\RktSym{label}\mbox{\hphantom{\Scribtexttt{x}}}\RktVal{"Counter"}\RktPn{]}\RktPn{)}\RktPn{)}

\RktPn{(}\RktSym{define}\mbox{\hphantom{\Scribtexttt{x}}}\RktSym{container}

\mbox{\hphantom{\Scribtexttt{xx}}}\RktPn{(}\RktSym{new}\mbox{\hphantom{\Scribtexttt{x}}}\RktSym{horizontal{-}panel\%}\mbox{\hphantom{\Scribtexttt{x}}}\RktPn{[}\RktSym{parent}\mbox{\hphantom{\Scribtexttt{x}}}\RktSym{f}\RktPn{]}\RktPn{)}\RktPn{)}

\RktPn{(}\RktSym{define}\mbox{\hphantom{\Scribtexttt{x}}}\RktSym{count}\mbox{\hphantom{\Scribtexttt{x}}}\RktVal{0}\RktPn{)}

\RktPn{(}\RktSym{define}\mbox{\hphantom{\Scribtexttt{x}}}\RktPn{(}\RktSym{update{-}count}\mbox{\hphantom{\Scribtexttt{x}}}\RktSym{f}\RktPn{)}

\mbox{\hphantom{\Scribtexttt{xx}}}\RktPn{(}\RktSym{set{\hbox{\texttt{!}}}}\mbox{\hphantom{\Scribtexttt{x}}}\RktSym{count}\mbox{\hphantom{\Scribtexttt{x}}}\RktPn{(}\RktSym{f}\mbox{\hphantom{\Scribtexttt{x}}}\RktSym{count}\RktPn{)}\RktPn{)}

\mbox{\hphantom{\Scribtexttt{xx}}}\RktPn{(}\RktSym{define}\mbox{\hphantom{\Scribtexttt{x}}}\RktSym{new{-}label}\mbox{\hphantom{\Scribtexttt{x}}}\RktPn{(}\RktSym{number{-}{\Stttextmore}string}\mbox{\hphantom{\Scribtexttt{x}}}\RktSym{count}\RktPn{)}\RktPn{)}

\mbox{\hphantom{\Scribtexttt{xx}}}\RktPn{(}\RktSym{send}\mbox{\hphantom{\Scribtexttt{x}}}\RktSym{count{-}label}\mbox{\hphantom{\Scribtexttt{x}}}\RktSym{set{-}label}\mbox{\hphantom{\Scribtexttt{x}}}\RktSym{new{-}label}\RktPn{)}\RktPn{)}

\RktPn{(}\RktSym{define}\mbox{\hphantom{\Scribtexttt{x}}}\RktSym{minus{-}button}

\mbox{\hphantom{\Scribtexttt{xx}}}\RktPn{(}\RktSym{new}\mbox{\hphantom{\Scribtexttt{x}}}\RktSym{button\%}\mbox{\hphantom{\Scribtexttt{x}}}\RktPn{[}\RktSym{parent}\mbox{\hphantom{\Scribtexttt{x}}}\RktSym{container}\RktPn{]}

\mbox{\hphantom{\Scribtexttt{xxxxxxx}}}\RktPn{[}\RktSym{label}\mbox{\hphantom{\Scribtexttt{x}}}\RktVal{"{-}"}\RktPn{]}

\mbox{\hphantom{\Scribtexttt{xxxxxxx}}}\RktPn{[}\RktSym{callback}\mbox{\hphantom{\Scribtexttt{x}}}\RktPn{(}\RktSym{$\lambda$}\mbox{\hphantom{\Scribtexttt{x}}}\RktSym{{\char`\_}}\mbox{\hphantom{\Scribtexttt{x}}}\RktPn{(}\RktSym{update{-}count}\mbox{\hphantom{\Scribtexttt{x}}}\RktSym{sub1}\RktPn{)}\RktPn{)}\RktPn{]}\RktPn{)}\RktPn{)}

\RktPn{(}\RktSym{define}\mbox{\hphantom{\Scribtexttt{x}}}\RktSym{count{-}label}

\mbox{\hphantom{\Scribtexttt{xx}}}\RktPn{(}\RktSym{new}\mbox{\hphantom{\Scribtexttt{x}}}\RktSym{message\%}\mbox{\hphantom{\Scribtexttt{x}}}\RktPn{[}\RktSym{parent}\mbox{\hphantom{\Scribtexttt{x}}}\RktSym{container}\RktPn{]}

\mbox{\hphantom{\Scribtexttt{xxxxxxx}}}\RktPn{[}\RktSym{label}\mbox{\hphantom{\Scribtexttt{x}}}\RktVal{"0"}\RktPn{]}

\mbox{\hphantom{\Scribtexttt{xxxxxxx}}}\RktPn{[}\RktSym{auto{-}resize}\mbox{\hphantom{\Scribtexttt{x}}}\RktVal{\#t}\RktPn{]}\RktPn{)}\RktPn{)}

\RktPn{(}\RktSym{define}\mbox{\hphantom{\Scribtexttt{x}}}\RktSym{plus{-}button}

\mbox{\hphantom{\Scribtexttt{xx}}}\RktPn{(}\RktSym{new}\mbox{\hphantom{\Scribtexttt{x}}}\RktSym{button\%}\mbox{\hphantom{\Scribtexttt{x}}}\RktPn{[}\RktSym{parent}\mbox{\hphantom{\Scribtexttt{x}}}\RktSym{container}\RktPn{]}

\mbox{\hphantom{\Scribtexttt{xxxxxxx}}}\RktPn{[}\RktSym{label}\mbox{\hphantom{\Scribtexttt{x}}}\RktVal{"+"}\RktPn{]}

\mbox{\hphantom{\Scribtexttt{xxxxxxx}}}\RktPn{[}\RktSym{callback}\mbox{\hphantom{\Scribtexttt{x}}}\RktPn{(}\RktSym{$\lambda$}\mbox{\hphantom{\Scribtexttt{x}}}\RktSym{{\char`\_}}\mbox{\hphantom{\Scribtexttt{x}}}\RktPn{(}\RktSym{update{-}count}\mbox{\hphantom{\Scribtexttt{x}}}\RktSym{add1}\RktPn{)}\RktPn{)}\RktPn{]}\RktPn{)}\RktPn{)}

\RktPn{(}\RktSym{send}\mbox{\hphantom{\Scribtexttt{x}}}\RktSym{f}\mbox{\hphantom{\Scribtexttt{x}}}\RktSym{show}\mbox{\hphantom{\Scribtexttt{x}}}\RktVal{\#t}\RktPn{)}\end{SingleColumn}\end{RktBlk}\end{FigureInside}\end{Centerfigure}

\Centertext{\Legend{\FigureTarget{\label{t:x28counter_x28x22figurex22_x22oopx2dcounterx2erktx22x29x29}\textbf{Figure}~\textbf{1}. }{t:x28counter_x28x22figurex22_x22oopx2dcounterx2erktx22x29x29}A counter GUI using Racket GUI{\textquotesingle}s object{-}oriented widgets.}}\end{Figure}

Figure~\hyperref[t:x28counter_x28x22figurex22_x22oopx2dcounterx2erktx22x29x29]{\FigureRef{1}{t:x28counter_x28x22figurex22_x22oopx2dcounterx2erktx22x29x29}} demonstrates typical Racket GUI code: it
renders a counter with buttons to increment and decrement a number.
First, we create a top{-}level window container, called a \RktSym{frame\%}.
To lay out the controls horizontally, we nest a
\RktSym{horizontal{-}panel\%} as a child of the window. We define the count
state and a procedure to simultaneously update the count and its
associated label. Next, we create the buttons and label for the counter.
Lastly, we call the \RktSym{show} method on the \RktSym{frame\%} to
render it for the user.

The code in figure~\hyperref[t:x28counter_x28x22figurex22_x22oopx2dcounterx2erktx22x29x29]{\FigureRef{1}{t:x28counter_x28x22figurex22_x22oopx2dcounterx2erktx22x29x29}} has several shortcomings. It is
verbose relative to the complexity of the GUI it describes and organized
in a way that obscures the structure of the resulting interface. The
programmer manually synchronizes application state, like the count, and
UI state, like the message label, by mutation.

\begin{Figure}\begin{Centerfigure}\begin{FigureInside}\begin{RktBlk}\begin{SingleColumn}\RktModLink{\RktMod{\#lang}}\mbox{\hphantom{\Scribtexttt{x}}}\RktModLink{\RktSym{racket/gui/easy}}

\RktPn{(}\RktSym{define}\mbox{\hphantom{\Scribtexttt{x}}}\RktSym{@count}\mbox{\hphantom{\Scribtexttt{x}}}\RktPn{(}\RktSym{@}\mbox{\hphantom{\Scribtexttt{x}}}\RktVal{0}\RktPn{)}\RktPn{)}

\RktPn{(}\RktSym{render}

\mbox{\hphantom{\Scribtexttt{x}}}\RktPn{(}\RktSym{window}

\mbox{\hphantom{\Scribtexttt{xx}}}\RktPn{\#{\hbox{\texttt{:}}}title}\mbox{\hphantom{\Scribtexttt{x}}}\RktVal{"Counter"}

\mbox{\hphantom{\Scribtexttt{xx}}}\RktPn{(}\RktSym{hpanel}

\mbox{\hphantom{\Scribtexttt{xxx}}}\RktPn{(}\RktSym{button}\mbox{\hphantom{\Scribtexttt{x}}}\RktVal{"{-}"}\mbox{\hphantom{\Scribtexttt{x}}}\RktPn{(}\RktSym{$\lambda$}\mbox{\hphantom{\Scribtexttt{x}}}\RktPn{(}\RktPn{)}\mbox{\hphantom{\Scribtexttt{x}}}\RktPn{(}\RktSym{{\Stttextless}\textasciitilde{}}\mbox{\hphantom{\Scribtexttt{x}}}\RktSym{@count}\mbox{\hphantom{\Scribtexttt{x}}}\RktSym{sub1}\RktPn{)}\RktPn{)}\RktPn{)}

\mbox{\hphantom{\Scribtexttt{xxx}}}\RktPn{(}\RktSym{text}\mbox{\hphantom{\Scribtexttt{x}}}\RktPn{(}\RktSym{\textasciitilde{}{\Stttextmore}}\mbox{\hphantom{\Scribtexttt{x}}}\RktSym{@count}\mbox{\hphantom{\Scribtexttt{x}}}\RktSym{number{-}{\Stttextmore}string}\RktPn{)}\RktPn{)}

\mbox{\hphantom{\Scribtexttt{xxx}}}\RktPn{(}\RktSym{button}\mbox{\hphantom{\Scribtexttt{x}}}\RktVal{"+"}\mbox{\hphantom{\Scribtexttt{x}}}\RktPn{(}\RktSym{$\lambda$}\mbox{\hphantom{\Scribtexttt{x}}}\RktPn{(}\RktPn{)}\mbox{\hphantom{\Scribtexttt{x}}}\RktPn{(}\RktSym{{\Stttextless}\textasciitilde{}}\mbox{\hphantom{\Scribtexttt{x}}}\RktSym{@count}\mbox{\hphantom{\Scribtexttt{x}}}\RktSym{add1}\RktPn{)}\RktPn{)}\RktPn{)}\RktPn{)}\RktPn{)}\RktPn{)}\end{SingleColumn}\end{RktBlk}\end{FigureInside}\end{Centerfigure}

\Centertext{\Legend{\FigureTarget{\label{t:x28counter_x28x22figurex22_x22easyx2dcounterx2erktx22x29x29}\textbf{Figure}~\textbf{2}. }{t:x28counter_x28x22figurex22_x22easyx2dcounterx2erktx22x29x29}A counter GUI using GUI Easy{\textquotesingle}s functional widgets.}}\end{Figure}

\begin{Figure}\begin{Centerfigure}\begin{FigureInside}\includegraphics[scale=0.5]{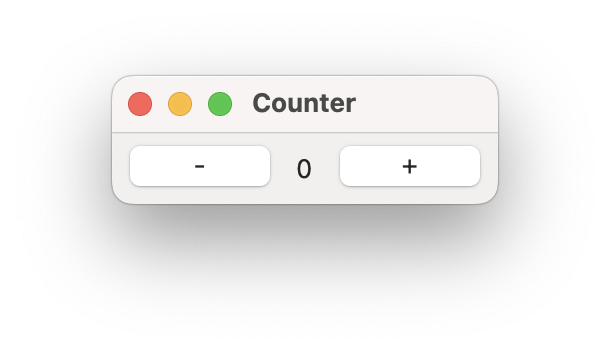}\end{FigureInside}\end{Centerfigure}

\Centertext{\Legend{\FigureTarget{\label{t:x28counter_x28x22figurex22_x22screenshotx2dcounterx2epngx22x29x29}\textbf{Figure}~\textbf{3}. }{t:x28counter_x28x22figurex22_x22screenshotx2dcounterx2epngx22x29x29}The rendered counter GUI on macOS.}}\end{Figure}

GUI Easy is a functional shell for Racket{'}s GUI system based on
observable values and function composition that aims to solve the
problems with the imperative object{-}based APIs\Autobibref{~[\hyperref[t:x28autobib_x22Bogdan_PopaAnnouncing_GUI_Easy2021httpsx3ax2fx2fdefnx2eiox2f2021x2f08x2f01x2fannx2dguix2deasyx2fx22x29]{\AutobibLink{26}}]}. You
can install the package \Scribtexttt{gui{-}easy} through the menu of
DrRacket\Autobibref{~[\hyperref[t:x28autobib_x22Robert_Bruce_Findlerx2c_John_Clementsx2c_Cormac_Flanaganx2c_Matthew_Flattx2c_Shriram_Krishnamurthix2c_Paul_Stecklerx2c_and_Matthias_FelleisenDrSchemex3a_A_programming_environment_for_Schemex2eJournal_of_Functional_Programming_12x282x29x2c_ppx2e_159x2dx2d1822002x22x29]{\AutobibLink{11}}]} or using the Racket command{-}line tool
\Scribtexttt{raco}\Autobibref{~[\hyperref[t:x28autobib_x22racox3a_Racket_Commandx2dLine_Tools2010httpsx3ax2fx2fdocsx2eracketx2dlangx2eorgx2fracox2findexx2ehtmlx22x29]{\AutobibLink{27}}]}, after which you can run the examples in
DrRacket or from your favorite programming environment.

With GUI Easy, the code in figure~\hyperref[t:x28counter_x28x22figurex22_x22easyx2dcounterx2erktx22x29x29]{\FigureRef{2}{t:x28counter_x28x22figurex22_x22easyx2dcounterx2erktx22x29x29}} resolves the
previous shortcomings. As state, we define an observable
\RktSym{@count} whose initial value is the number \RktVal{0}. Then, we
\RktSym{render} an interface composed of widgets like \RktSym{window},
\RktSym{hpanel}, \RktSym{button}, and \RktSym{text}. Widget properties,
such as size or label, may be constant values or observables. The
rendered widgets automatically update when their observable inputs
change, similar to systems like React\Autobibref{~[\hyperref[t:x28autobib_x22Meta_Open_SourceReact2023httpsx3ax2fx2freactx2edevRetrieved_June_2023x2ex22x29]{\AutobibLink{29}}]} and
SwiftUI\Autobibref{~[\hyperref[t:x28autobib_x22AppleSwiftUI2023httpsx3ax2fx2fdeveloperx2eapplex2ecomx2fxcodex2fswiftuix2fRetrieved_June_2023x2ex22x29]{\AutobibLink{1}}]}. In this example, pressing the buttons causes
the counter to be updated, which updates the text label.

In this report, we examine the difficulties of programming with
object{-}oriented GUI systems and motivate the search for a different
system in \ChapRef{\SectionNumberLink{t:x28part_x22Ax5fTalex5fofx5fTwox5fProgrammersx22x29}{2}}{A Tale of Two Programmers}, describe the main GUI Easy
abstractions in \ChapRef{\SectionNumberLink{t:x28part_x22GUIx5fEasyx5fOverviewx22x29}{3}}{GUI Easy Overview}, report on our experience
constructing large GUI programs in \ChapRef{\SectionNumberLink{t:x28part_x22archx2dfrostx22x29}{4}}{The Architecture of Frosthaven}, explore key
architectural lessons in \ChapRef{\SectionNumberLink{t:x28part_x22Architecturalx5fLessonsx22x29}{5}}{Architectural Lessons}, and explore
related trends in GUI programming in \ChapRef{\SectionNumberLink{t:x28part_x22relatedx5fworkx22x29}{6}}{Related Work}.

\sectionNewpage

\Ssection{A Tale of Two Programmers}{A Tale of Two Programmers}\label{t:x28part_x22Ax5fTalex5fofx5fTwox5fProgrammersx22x29}

We present the origin stories for two projects. First, Bogdan describes
his frustrations with Racket{'}s GUI system that drove him to create GUI
Easy. Second, Ben describes his desire to construct a large GUI program
using a functional approach. The happy union of these two desires taught
us the architectural lessons we present in \ChapRef{\SectionNumberLink{t:x28part_x22archx2dfrostx22x29}{4}}{The Architecture of Frosthaven}.

\Ssubsection{Quest for Easier GUIs}{Quest for Easier GUIs}\label{t:x28part_x22questx2dforx2dguix2deasyx22x29}

Bogdan{'}s day job involved writing many small GUI tools for internal use.
The Racket GUI framework proved an excellent way to build those types of
tools as it provides fast iteration times, portability across major
operating systems, and distribution of self{-}contained applications.

Over time, however, Bogdan was repeatedly annoyed by the same
inconveniences. Racket{'}s class system requires verbose code. Each
project manages state in its own way. Racket GUI{'}s primary means of
constructing view hierarchies is to construct child widgets with
references to their parent widgets, which makes composition especially
frustrating since individual components must always be parameterized
over their parent.

Since Racket GUI offers no special support for managing application
state, Bogdan had to bring his own state management to the
table, leading to ad hoc solutions for every new project. See
\RktSym{update{-}count} in figure~\hyperref[t:x28counter_x28x22figurex22_x22oopx2dcounterx2erktx22x29x29]{\FigureRef{1}{t:x28counter_x28x22figurex22_x22oopx2dcounterx2erktx22x29x29}} for an example
of ad hoc state management. This motivated the observable abstraction
in GUI Easy. In the next section, we will see how observables and
observable{-}aware views combine to automatically connect GUI widgets and
state changes.

Bogdan found it inconvenient that constructing most widgets requires a
reference to a parent widget. Consider the following piece of Racket
code:

\begin{SCodeFlow}\begin{RktBlk}\begin{SingleColumn}\RktPn{(}\RktSym{define}\mbox{\hphantom{\Scribtexttt{x}}}\RktSym{f}\mbox{\hphantom{\Scribtexttt{x}}}\RktPn{(}\RktSym{new}\mbox{\hphantom{\Scribtexttt{x}}}\RktSym{frame\%}\mbox{\hphantom{\Scribtexttt{x}}}\RktPn{[}\RktSym{label}\mbox{\hphantom{\Scribtexttt{x}}}\RktVal{"A window"}\RktPn{]}\RktPn{)}\RktPn{)}

\RktPn{(}\RktSym{define}\mbox{\hphantom{\Scribtexttt{x}}}\RktSym{msg}

\mbox{\hphantom{\Scribtexttt{xx}}}\RktPn{(}\RktSym{new}\mbox{\hphantom{\Scribtexttt{x}}}\RktSym{message\%}\mbox{\hphantom{\Scribtexttt{x}}}\RktPn{[}\RktSym{parent}\mbox{\hphantom{\Scribtexttt{x}}}\RktSym{f}\RktPn{]}

\mbox{\hphantom{\Scribtexttt{xxxxxxx}}}\RktPn{[}\RktSym{label}\mbox{\hphantom{\Scribtexttt{x}}}\RktVal{"Hello World"}\RktPn{]}\RktPn{)}\RktPn{)}\end{SingleColumn}\end{RktBlk}\end{SCodeFlow}

We cannot create the message object before the frame object in this
case, since we need a \RktSym{parent} for the message object. This
constrains how we can organize code. To work around the issue, we
can abstract over message object construction, but that needlessly
complicates wiring up interfaces. This motivated Bogdan to come up with
the view abstraction in GUI Easy. In \ChapRef{\SectionNumberLink{t:x28part_x22GUIx5fEasyx5fOverviewx22x29}{3}}{GUI Easy Overview}, we will
see how views permit functional abstraction, enabling new organizational
approaches that we will explore in \ChapRef{\SectionNumberLink{t:x28part_x22archx2dfrostx22x29}{4}}{The Architecture of Frosthaven}.

\Ssubsection{Embarking for the Town of Frosthaven}{Embarking for the Town of Frosthaven}\label{t:x28part_x22embarkingx22x29}

Ben enjoys boardgames with a group of friends, especially
Frosthaven\Autobibref{~[\hyperref[t:x28autobib_x22Cephalofair_GamesFrosthaven2023httpsx3ax2fx2fcephalofairx2ecomx2fpagesx2ffrosthavenx22x29]{\AutobibLink{4}}]}, the sequel to Gloomhaven. Due to its
highly complex nature, Frosthaven includes lots of tokens, cards, and
other physical pieces that the players must manipulate to play the game.
This includes tracking monsters{'} health and conditions, the strength
of six magical elements that power special abilities, and more. The
original Gloomhaven game had a helper application for mobile devices to
reduce physical manipulation; at one point, it appeared Frosthaven would
not receive the same treatment.

Ben, a programmer, decided to solve the problem for his personal gaming
group by creating his own helper application. But how? Having never
created a complex GUI program, Ben was intimidated by classic
object{-}oriented systems like Racket{'}s GUI toolkit. To a programmer with
intimate knowledge of the class, method, and event relationships, such a
system may feel natural. To the novice, GUI Easy represents a simpler,
functional, path to interface programming.

GUI Easy makes it possible to build a complex system out of simple
parts: functions and data. Ben was familiar with functional programming
and grokked GUI Easy, so he started programming the Frosthaven
Manager\Autobibref{~[\hyperref[t:x28autobib_x22Dx2e_Ben_Knoblefrosthavenx2dmanager2022httpsx3ax2fx2fgithubx2ecomx2fbenknoblex2ffrosthavenx2dmanagerx22x29]{\AutobibLink{20}}]} with GUI Easy in 2022.

\sectionNewpage

\Ssection{GUI Easy Overview}{GUI Easy Overview}\label{t:x28part_x22GUIx5fEasyx5fOverviewx22x29}

The goal of GUI Easy is to simplify user interface construction in
Racket by wrapping its imperative API in a functional shell. GUI Easy
can be broadly split up into two parts: \textit{observables} and
\textit{views}.

Observables contain values and notify subscribed observers of changes
to their contents. \SecRefUC{\SectionNumberLink{t:x28part_x22Observablex5fValuesx22x29}{3.1}}{Observable Values} explains the observable
operators.

Views are representations of Racket GUI widget trees that, when
rendered, produce concrete instances of those trees and handle the
details of wiring state and widgets together. We discuss the view
abstraction in more detail in \SecRef{\SectionNumberLink{t:x28part_x22viewx5fdetailx22x29}{3.2}}{Views as Functions}.

The core abstractions of observables and views correspond to a
model{-}view{-}controller (MVC) architecture for graphical applications, as
popularized by Smalltalk{-}80\Autobibref{~[\hyperref[t:x28autobib_x22Adele_GoldbergSmalltalkx2d80x3a_The_Interactive_Programming_EnvironmentAddisonx2dWesley_Publishersx2e1983x22x29]{\AutobibLink{15}}, \hyperref[t:x28autobib_x22Glenn_Ex2e_Krasner_and_Stephen_Tx2e_PopeA_description_of_the_modelx2dviewx2dcontroller_user_interface_paradigm_in_the_Smalltalkx2d80_systemJournal_of_Objectx2dOriented_Programming_1x283x29x2c_ppx2e_26x2dx2d491988httpsx3ax2fx2fwwwx2eicsx2eucix2eedux2fx7eredmilesx2fics227x2dSQ04x2fpapersx2fKrasnerPope88x2epdfx22x29]{\AutobibLink{21}}]}. We describe the
correspondence in \SecRef{\SectionNumberLink{t:x28part_x22mvcx22x29}{3.3}}{Models, Views, and Controllers}.

\begin{Figure}\begin{Centerfigure}\begin{FigureInside}\begin{RktBlk}\begin{SingleColumn}\RktModLink{\RktMod{\#lang}}\mbox{\hphantom{\Scribtexttt{x}}}\RktModLink{\RktSym{racket/gui/easy}}

\RktPn{(}\RktSym{define}\mbox{\hphantom{\Scribtexttt{x}}}\RktPn{(}\RktSym{counter}\mbox{\hphantom{\Scribtexttt{x}}}\RktSym{@count}\mbox{\hphantom{\Scribtexttt{x}}}\RktSym{action}\RktPn{)}

\mbox{\hphantom{\Scribtexttt{xx}}}\RktPn{(}\RktSym{hpanel}

\mbox{\hphantom{\Scribtexttt{xxx}}}\RktPn{(}\RktSym{button}\mbox{\hphantom{\Scribtexttt{x}}}\RktVal{"{-}"}\mbox{\hphantom{\Scribtexttt{x}}}\RktPn{(}\RktSym{$\lambda$}\mbox{\hphantom{\Scribtexttt{x}}}\RktPn{(}\RktPn{)}\mbox{\hphantom{\Scribtexttt{x}}}\RktPn{(}\RktSym{action}\mbox{\hphantom{\Scribtexttt{x}}}\RktSym{sub1}\RktPn{)}\RktPn{)}\RktPn{)}

\mbox{\hphantom{\Scribtexttt{xxx}}}\RktPn{(}\RktSym{text}\mbox{\hphantom{\Scribtexttt{x}}}\RktPn{(}\RktSym{\textasciitilde{}{\Stttextmore}}\mbox{\hphantom{\Scribtexttt{x}}}\RktSym{@count}\mbox{\hphantom{\Scribtexttt{x}}}\RktSym{number{-}{\Stttextmore}string}\RktPn{)}\RktPn{)}

\mbox{\hphantom{\Scribtexttt{xxx}}}\RktPn{(}\RktSym{button}\mbox{\hphantom{\Scribtexttt{x}}}\RktVal{"+"}\mbox{\hphantom{\Scribtexttt{x}}}\RktPn{(}\RktSym{$\lambda$}\mbox{\hphantom{\Scribtexttt{x}}}\RktPn{(}\RktPn{)}\mbox{\hphantom{\Scribtexttt{x}}}\RktPn{(}\RktSym{action}\mbox{\hphantom{\Scribtexttt{x}}}\RktSym{add1}\RktPn{)}\RktPn{)}\RktPn{)}\RktPn{)}\RktPn{)}

\mbox{\hphantom{\Scribtexttt{x}}}

\RktPn{(}\RktSym{define}\mbox{\hphantom{\Scribtexttt{x}}}\RktSym{@c1}\mbox{\hphantom{\Scribtexttt{x}}}\RktPn{(}\RktSym{@}\mbox{\hphantom{\Scribtexttt{x}}}\RktVal{0}\RktPn{)}\RktPn{)}

\RktPn{(}\RktSym{define}\mbox{\hphantom{\Scribtexttt{x}}}\RktSym{@c2}\mbox{\hphantom{\Scribtexttt{x}}}\RktPn{(}\RktSym{@}\mbox{\hphantom{\Scribtexttt{x}}}\RktVal{5}\RktPn{)}\RktPn{)}

\mbox{\hphantom{\Scribtexttt{x}}}

\RktPn{(}\RktSym{render}

\mbox{\hphantom{\Scribtexttt{x}}}\RktPn{(}\RktSym{window}

\mbox{\hphantom{\Scribtexttt{xx}}}\RktPn{\#{\hbox{\texttt{:}}}title}\mbox{\hphantom{\Scribtexttt{x}}}\RktVal{"Counters"}

\mbox{\hphantom{\Scribtexttt{xx}}}\RktPn{(}\RktSym{counter}\mbox{\hphantom{\Scribtexttt{x}}}\RktSym{@c1}\mbox{\hphantom{\Scribtexttt{x}}}\RktPn{(}\RktSym{$\lambda$}\mbox{\hphantom{\Scribtexttt{x}}}\RktPn{(}\RktSym{proc}\RktPn{)}\mbox{\hphantom{\Scribtexttt{x}}}\RktPn{(}\RktSym{{\Stttextless}\textasciitilde{}}\mbox{\hphantom{\Scribtexttt{x}}}\RktSym{@c1}\mbox{\hphantom{\Scribtexttt{x}}}\RktSym{proc}\RktPn{)}\RktPn{)}\RktPn{)}

\mbox{\hphantom{\Scribtexttt{xx}}}\RktPn{(}\RktSym{counter}\mbox{\hphantom{\Scribtexttt{x}}}\RktSym{@c2}\mbox{\hphantom{\Scribtexttt{x}}}\RktPn{(}\RktSym{$\lambda$}\mbox{\hphantom{\Scribtexttt{x}}}\RktPn{(}\RktSym{proc}\RktPn{)}\mbox{\hphantom{\Scribtexttt{x}}}\RktPn{(}\RktSym{{\Stttextless}\textasciitilde{}}\mbox{\hphantom{\Scribtexttt{x}}}\RktSym{@c2}\mbox{\hphantom{\Scribtexttt{x}}}\RktSym{proc}\RktPn{)}\RktPn{)}\RktPn{)}\RktPn{)}\RktPn{)}\end{SingleColumn}\end{RktBlk}\end{FigureInside}\end{Centerfigure}

\Centertext{\Legend{\FigureTarget{\label{t:x28counter_x28x22figurex22_x22easyx2dcounterx2dreusex2erktx22x29x29}\textbf{Figure}~\textbf{4}. }{t:x28counter_x28x22figurex22_x22easyx2dcounterx2dreusex2erktx22x29x29}Component re{-}use in GUI Easy. Multiple counter widgets can be created from a single definition.}}\end{Figure}

\Ssubsection{Observable Values}{Observable Values}\label{t:x28part_x22Observablex5fValuesx22x29}

The core of the observable abstraction is that arbitrary observers
can react to changes in the contents of an observable. Application
developers programming with GUI Easy use a few core operators to
construct and manipulate observables.

We create observables with \RktSym{@}. By convention, we prefix
observable bindings with the same sigil.

We can change the contents of an observable using \RktSym{{\Stttextless}\textasciitilde{}}. This
procedure takes as arguments an observable and a procedure of one
argument, representing the current value, to generate a new value. Every
change is propagated to any observers registered at the time of the
update.

We can derive new observables from existing ones using \RktSym{\textasciitilde{}{\Stttextmore}}.
This procedure takes an observable and a procedure of one argument, the
current value. A derived observable changes with the observable it is
derived from by applying its mapping procedure to the values of its
input observable. In figure~\hyperref[t:x28counter_x28x22figurex22_x22easyx2dcounterx2dreusex2erktx22x29x29]{\FigureRef{4}{t:x28counter_x28x22figurex22_x22easyx2dcounterx2dreusex2erktx22x29x29}}, the derived
observable \RktPn{(}\RktSym{\textasciitilde{}{\Stttextmore}}\Scribtexttt{ }\RktSym{@count}\Scribtexttt{ }\RktSym{number{-}{\Stttextmore}string}\RktPn{)} changes every time
\RktSym{@count} is updated by \RktSym{{\Stttextless}\textasciitilde{}}; its value is the result of
applying \RktSym{number{-}{\Stttextmore}string} to the value of \RktSym{@count}. We
cannot directly update derived observables.

We can peek at an observable with \RktSym{obs{-}peek}, which returns
the contents of the observable. This operation is useful to get
point{-}in{-}time values out of observables when displaying modal dialogs or
other views that require a snapshot of the state.

\Ssubsection{Views as Functions}{Views as Functions}\label{t:x28part_x22viewx5fdetailx22x29}

Views are functions that return a \RktSym{view{\Stttextless}\%{\Stttextmore}} instance, whose
underlying details we will cover in \SecRef{\SectionNumberLink{t:x28part_x22viewx5fimplx22x29}{5.2}}{\Scribtexttt{view{\Stttextless}\%{\Stttextmore}}: Functional Shell, Imperative Core}. Views might wrap
a specific GUI widget, like a text message or button, or they might
construct a tree of smaller views, forming a larger component. Both are
synonymous with {``}view{''} in this report. We have already seen many examples
of views like \RktSym{text}, \RktSym{hpanel}, and \RktSym{counter}.

Views are typically observable{-}aware in ways that make sense for each
individual view. For instance, the \RktSym{text} view takes as input an
observable string and the rendered text label updates with changes to
that observable. Figure~\hyperref[t:x28counter_x28x22figurex22_x22easyx2dcounterx2dreusex2erktx22x29x29]{\FigureRef{4}{t:x28counter_x28x22figurex22_x22easyx2dcounterx2dreusex2erktx22x29x29}} shows an example of
a reusable counter component made by composing views together.

Many Racket GUI widgets are already wrapped by GUI Easy, but programmers
can implement the \RktSym{view{\Stttextless}\%{\Stttextmore}} interface themselves in order to
integrate arbitrary widgets, such as those from 3rd{-}party packages in
the Racket ecosystem, into their projects.

\Ssubsection{Models, Views, and Controllers}{Models, Views, and Controllers}\label{t:x28part_x22mvcx22x29}

The popular MVC architecture for graphical applications divides program
modules into models of the application domain, views of the models, and
controllers coupled with the views to translate user interactions into
commands that affect the model\Autobibref{~[\hyperref[t:x28autobib_x22Glenn_Ex2e_Krasner_and_Stephen_Tx2e_PopeA_description_of_the_modelx2dviewx2dcontroller_user_interface_paradigm_in_the_Smalltalkx2d80_systemJournal_of_Objectx2dOriented_Programming_1x283x29x2c_ppx2e_26x2dx2d491988httpsx3ax2fx2fwwwx2eicsx2eucix2eedux2fx7eredmilesx2fics227x2dSQ04x2fpapersx2fKrasnerPope88x2epdfx22x29]{\AutobibLink{21}}]}.

Racket GUI applications can be organized according to the MVC
architecture. In figure~\hyperref[t:x28counter_x28x22figurex22_x22oopx2dcounterx2erktx22x29x29]{\FigureRef{1}{t:x28counter_x28x22figurex22_x22oopx2dcounterx2erktx22x29x29}}, the model is an integer
\RktSym{count}; the view is the combination of \RktSym{button\%} and
\RktSym{mesage\%} objects, and the controller is the
\RktSym{update{-}count} procedure. Notice, however, that
explictly grouping the view objects into a single reusable component
requires contorting the code responsible for object creation. There is
no explicit support for the MVC pattern, though it can be used
implicitly.

GUI Easy encourages an MVC{-}like architecture through the observable and
view abstractions. Consider as an example
figure~\hyperref[t:x28counter_x28x22figurex22_x22easyx2dcounterx2dreusex2erktx22x29x29]{\FigureRef{4}{t:x28counter_x28x22figurex22_x22easyx2dcounterx2dreusex2erktx22x29x29}}: the observables \RktSym{@c1} and
\RktSym{@c2} form the model, and each is distinguished from ordinary
values. Similarly, the \RktSym{counter} procedure is both a GUI Easy
view and an MVC view. Finally, the controller{'}s role is fulfilled by the
\RktSym{action} callback, which gives the \RktSym{counter} consumer
control over how user interactions are translated to model updates.

In summary, the MVC architecture encouraged by GUI Easy uses observables
for models, GUI Easy views for views, and callbacks for controllers.

\sectionNewpage

\Ssection{The Architecture of Frosthaven}{The Architecture of Frosthaven}\label{t:x28part_x22archx2dfrostx22x29}

\begin{Figure}\begin{Centerfigure}\begin{FigureInside}\includegraphics[scale=0.18]{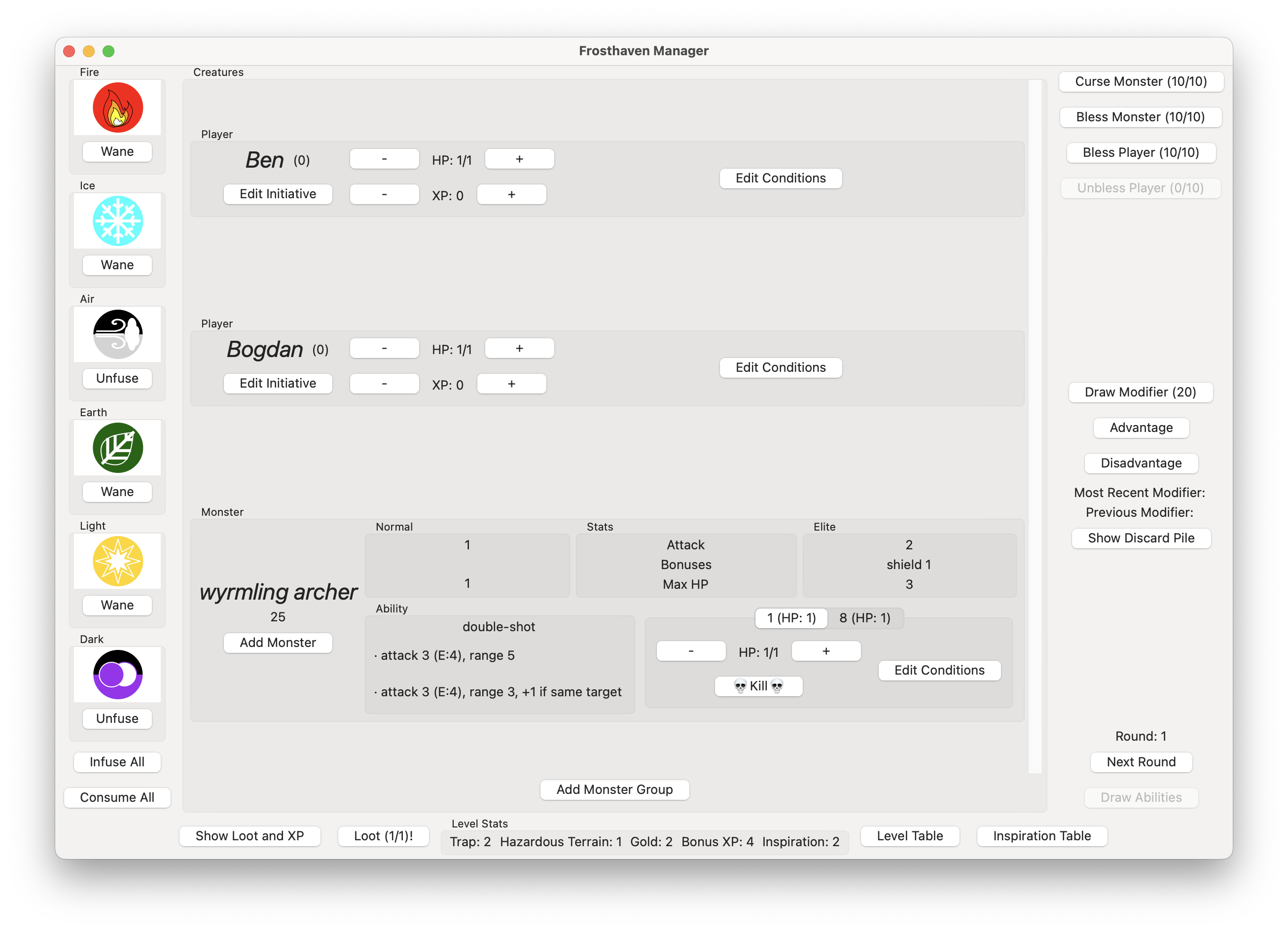}\end{FigureInside}\end{Centerfigure}

\Centertext{\Legend{\FigureTarget{\label{t:x28counter_x28x22figurex22_x22screenshotx2dfrosthavenx2epngx22x29x29}\textbf{Figure}~\textbf{5}. }{t:x28counter_x28x22figurex22_x22screenshotx2dfrosthavenx2epngx22x29x29}The Frosthaven Manager{\textquotesingle}s main window on macOS.}}\end{Figure}

In this section, we describe various pieces of a large GUI Easy
application, the Frosthaven Manager.

At time of writing, the Frosthaven Manager includes approximately
5000 lines of Racket code. About half of that code makes up the main
application by combining GUI Easy views with domain{-}specific code. Of
the remaining lines, approximately 1000 implement the data structures
and transformations responsible for the state of the game; 500 cover the
images it draws; 750 implement three user{-}programmable data{-}definition
languages\NoteBox{\NoteContent{\href{https://benknoble.github.io/frosthaven-manager/Programming_a_Scenario.html}{\Snolinkurl{https://benknoble.github.io/frosthaven-manager/Programming_a_Scenario.html}}}}; 300 test the project; the remaining lines are
small syntactic utilities. The Frosthaven Manager also has approximately
3000 lines of Scribble, a Racket prose and documentation language, which
includes a how{-}to{-}play guide and developer reference.

The Frosthaven Manager manipulates many kinds of data. This includes
game characters and their various attributes, monsters and their
attributes, randomized loot, the status of elemental effects, and more.
To organize and manipulate this data, Ben chose a {``}functional core,
imperative shell{''} architecture\Autobibref{~[\hyperref[t:x28autobib_x22Gary_BernhardtFunctional_Corex2c_Imperative_Shell2012httpsx3ax2fx2fwwwx2edestroyallsoftwarex2ecomx2fscreencastsx2fcatalogx2ffunctionalx2dcorex2dimperativex2dshellx22x29]{\AutobibLink{2}}]}.

The choice of a functional core and imperative shell has many benefits.
For example, core code is independent of the choice of UI presentation
and is independently testable or usable for other applications.
Functional cores also simplify programmer reasoning about application
data flow, keeping state change at the boundaries of the system.

In constructing the Frosthaven Manager, Ben organized the main data into
immutable records, enumerations, and collections alongside pure
functions that transform data according to the rules of the game. We
thus say that the Frosthaven Manager uses a functional core.

Layered atop the functional core we find two more major components in
the Frosthaven Manager: GUI{-}specific data and domain{-}specific views
built on GUI Easy. In many ways, Ben took the functional approach here,
too. GUI{-}related data is organized along typical idioms and paired with
transformation functions. Despite these functional qualities, since most
of the relevant data is observable or intended to be observable, the
resulting system feels far more imperative. For example, pure
transformations from the functional layers are paired with observable
updates{---}akin to mutations{---}for real effect on the state of the GUI.
As a result, though many important and reusable views seem pure, they
are easily combined into a highly imperative system. These views and
updates form the Frosthaven Manager{'}s imperative shell.

The Frosthaven Manager{'}s main GUI comprises many smaller reusable views.
By analogy with functional programming{'}s building
blocks{---}functions{---}small reusable views permit us to construct large
systems via composition. We will discuss the design principles behind
reusable views in \SecRef{\SectionNumberLink{t:x28part_x22Reusablex5fViewsx22x29}{5.1}}{Reusable Views}.

\sectionNewpage

\Ssection{Architectural Lessons}{Architectural Lessons}\label{t:x28part_x22Architecturalx5fLessonsx22x29}

In this section, we cover the major lessons learned while developing
these systems. First, reusable views (\SecRef{\SectionNumberLink{t:x28part_x22Reusablex5fViewsx22x29}{5.1}}{Reusable Views}) permit
interface composition akin to functional composition by constraining how
state is manipulated. Second, wrapping an imperative API with a
functional shell (\SecRef{\SectionNumberLink{t:x28part_x22viewx5fimplx22x29}{5.2}}{\Scribtexttt{view{\Stttextless}\%{\Stttextmore}}: Functional Shell, Imperative Core}) allows programmers to use
functional techniques and architectures when constructing imperative
systems. Third, inversion of control (\SecRef{\SectionNumberLink{t:x28part_x22iocx22x29}{5.3}}{Inversion of Control}) creates an
extensible application skeleton.

\Ssubsection{Reusable Views}{Reusable Views}\label{t:x28part_x22Reusablex5fViewsx22x29}

Our experience building applications taught us to prefer reusable views
where possible. Much like pure functions, a reusable view is composable
and is subject to constraints on state manipulation. All the views
provided by GUI Easy are reusable as described in this section.

There is one major design constraint on reusable views. \emph{Views
should not directly manipulate external state.} This is analogous to
the rule for pure functions, and all the same arguments apply to show
that manipulating external state makes a view less reusable. This leads
naturally to the principle {``}data down, actions up,{''} or \emph{DDAU}. It
also guides us to make decisions about which state to centralize at the
highest levels of the GUI and which state to localize in reusable views.

DDAU prescribes how a reusable view should manipulate state. The {``}data
down{''} prescription means that all necessary data must be inputs to a
view. Recall the \RktSym{counter} view from
figure~\hyperref[t:x28counter_x28x22figurex22_x22easyx2dcounterx2dreusex2erktx22x29x29]{\FigureRef{4}{t:x28counter_x28x22figurex22_x22easyx2dcounterx2dreusex2erktx22x29x29}}: the data needed to display the
value of the counter was an observable input to the view called
\RktSym{@count}. The {``}actions up{''} prescription means that views
should not directly manipulate state; instead, they should pass
actionable data back to their caller, which is better positioned to
decide how to manipulate state. Actions are represented by callbacks.
For the \RktSym{counter} view, the \RktSym{action} callback is passed a
procedure indicating whether the minus or plus button was clicked; the
caller of the \RktSym{counter} view decides how to react to user
manipulations of the GUI.

It would be generally unsafe to mutate observable inputs, as they
could be derived observables. Requiring informally that a particular
view{'}s observable inputs are not derived observables creates a trap
for programmers that want to reuse the view in novel contexts and
violates the principles of reusable views. Reusable views could take
separate input and output observable formal arguments to work around
this restriction, but that approach is generally less flexible and less
convenient for the user than callbacks.

Callbacks are also easier to compose than separate input and output
observables. For example, when a parent view uses a child view, it might
specify the child{'}s callback by wrapping the parent{'}s own callback. The
result is that events in the child are passed up through the parent,
with the parent able to intercept, modify, and filter events from the
child.

DDAU naturally bubbles application state up the layers of application
architecture, so that the top{-}level of an application contains all of
the necessary state. Callers pass the state, or a subset of it, down to
various component views and provide procedures to respond to actions.
This downward flow of state continues until we reach the bottom{-}most
layer. Sometimes, however, we need state that is neither the caller{'}s
nor callee{'}s responsibility. In such cases, a reusable view maintains
local state which it is free to manipulate. This is in keeping with the
tradition of optimizing functional programs by allowing interior{---}but
invisible{---}mutability.

The benefits of reusable views are threefold. Small reusable views are
amenable to independent testing. General{-}purpose views can be considered
for extraction to a separate library, much like generic data{-}structure
functions. Domain{-}specific views facilitate cohesion, such as visual
style for a GUI application.

While reusable views are a GUI{-}specific idea, the notions of DDAU and
constrained state management are also a more general lesson for
functional programming: identifying patterns of state manipulation and
constraining such state manipulation is a useful way to contain state in
a smaller portion of code and to permit functional techniques in the
remainder.

\Ssubsection{\Scribtexttt{view{\Stttextless}\%{\Stttextmore}}: Functional Shell, Imperative Core}{\Scribtexttt{view{\Stttextless}\%{\Stttextmore}}: Functional Shell, Imperative Core}\label{t:x28part_x22viewx5fimplx22x29}

The {``}Functional Core, Imperative Shell{''} architecture involves wrapping
a core of pure functional code with a shell of imperative commands. In
a twist on the paradigm, the core of GUI Easy views is an imperative
object lifecycle, while its shell is functional. In this section, we
describe that shell in detail and explain how it permits retaining
functional programming techniques when dealing with imperative systems.

The GUI object lifecycle is embodied by the \RktSym{view{\Stttextless}\%{\Stttextmore}} interface
(figure~\hyperref[t:x28counter_x28x22figurex22_x22viewx2difacex2erktx22x29x29]{\FigureRef{6}{t:x28counter_x28x22figurex22_x22viewx2difacex2erktx22x29x29}}). Implementations of the interface must
know how to \textit{create} widgets, how to \textit{update} them in
response to changed data dependencies, and how to \textit{destroy}
them if necessary\Autobibref{~[\hyperref[t:x28autobib_x22Bogdan_PopaAnnouncing_GUI_Easy2021httpsx3ax2fx2fdefnx2eiox2f2021x2f08x2f01x2fannx2dguix2deasyx2fx22x29]{\AutobibLink{26}}]}. They must also propagate data
dependencies up the object tree. Data dependencies are any observable
inputs to a view. The framework signals updates when dependencies
change, allowing \RktSym{view{\Stttextless}\%{\Stttextmore}}s to propagate updates to their wrapped
widgets. Crucially, \RktSym{view{\Stttextless}\%{\Stttextmore}} instances must be reusable, so they
must carefully associate any internal state they need with each rendered
widget.

\begin{Figure}\begin{Centerfigure}\begin{FigureInside}\begin{SCodeFlow}\begin{RktBlk}\begin{SingleColumn}\RktPn{(}\RktSym{define}\mbox{\hphantom{\Scribtexttt{x}}}\RktSym{view{\Stttextless}\%{\Stttextmore}}

\mbox{\hphantom{\Scribtexttt{xx}}}\RktPn{(}\RktSym{interface}\mbox{\hphantom{\Scribtexttt{x}}}\RktPn{(}\RktPn{)}

\mbox{\hphantom{\Scribtexttt{xxxx}}}\RktPn{[}\RktSym{dependencies}\mbox{\hphantom{\Scribtexttt{x}}}\RktPn{(}\RktSym{\mbox{{-}{\Stttextmore}}m}\mbox{\hphantom{\Scribtexttt{x}}}\RktPn{(}\RktSym{listof}\mbox{\hphantom{\Scribtexttt{x}}}\RktSym{obs{\hbox{\texttt{?}}}}\RktPn{)}\RktPn{)}\RktPn{]}

\mbox{\hphantom{\Scribtexttt{xxxx}}}\RktPn{[}\RktSym{create}\mbox{\hphantom{\Scribtexttt{x}}}\RktPn{(}\RktSym{\mbox{{-}{\Stttextmore}}m}\mbox{\hphantom{\Scribtexttt{x}}}\RktSym{container/c}\mbox{\hphantom{\Scribtexttt{x}}}\RktSym{widget/c}\RktPn{)}\RktPn{]}

\mbox{\hphantom{\Scribtexttt{xxxx}}}\RktPn{[}\RktSym{update}\mbox{\hphantom{\Scribtexttt{x}}}\RktPn{(}\RktSym{\mbox{{-}{\Stttextmore}}m}\mbox{\hphantom{\Scribtexttt{x}}}\RktSym{widget/c}\mbox{\hphantom{\Scribtexttt{x}}}\RktSym{obs{\hbox{\texttt{?}}}}\mbox{\hphantom{\Scribtexttt{x}}}\RktSym{any/c}\mbox{\hphantom{\Scribtexttt{x}}}\RktSym{void{\hbox{\texttt{?}}}}\RktPn{)}\RktPn{]}

\mbox{\hphantom{\Scribtexttt{xxxx}}}\RktPn{[}\RktSym{destroy}\mbox{\hphantom{\Scribtexttt{x}}}\RktPn{(}\RktSym{\mbox{{-}{\Stttextmore}}m}\mbox{\hphantom{\Scribtexttt{x}}}\RktSym{widget/c}\mbox{\hphantom{\Scribtexttt{x}}}\RktSym{void{\hbox{\texttt{?}}}}\RktPn{)}\RktPn{]}\RktPn{)}\RktPn{)}\end{SingleColumn}\end{RktBlk}\end{SCodeFlow}\end{FigureInside}\end{Centerfigure}

\Centertext{\Legend{\FigureTarget{\label{t:x28counter_x28x22figurex22_x22viewx2difacex2erktx22x29x29}\textbf{Figure}~\textbf{6}. }{t:x28counter_x28x22figurex22_x22viewx2difacex2erktx22x29x29}The \RktSym{view{\Stttextless}\%{\Stttextmore}} interface.}}\end{Figure}

\begin{Figure}\begin{Centerfigure}\begin{FigureInside}\begin{SCodeFlow}\begin{RktBlk}\begin{SingleColumn}\RktPn{(}\RktSym{require}\mbox{\hphantom{\Scribtexttt{x}}}\RktPn{(}\RktSym{prefix{-}in}\mbox{\hphantom{\Scribtexttt{x}}}\RktSym{gui{\hbox{\texttt{:}}}}\mbox{\hphantom{\Scribtexttt{x}}}\RktSym{racket/gui}\RktPn{)}\RktPn{)}

\RktPn{(}\RktSym{define}\mbox{\hphantom{\Scribtexttt{x}}}\RktSym{text\%}

\mbox{\hphantom{\Scribtexttt{xx}}}\RktPn{(}\RktSym{class*}\mbox{\hphantom{\Scribtexttt{x}}}\RktSym{object\%}\mbox{\hphantom{\Scribtexttt{x}}}\RktPn{(}\RktSym{view{\Stttextless}\%{\Stttextmore}}\RktPn{)}

\mbox{\hphantom{\Scribtexttt{xxxx}}}\RktPn{(}\RktSym{init{-}field}\mbox{\hphantom{\Scribtexttt{x}}}\RktSym{@label}\RktPn{)}\mbox{\hphantom{\Scribtexttt{x}}}\RktPn{(}\RktSym{super{-}new}\RktPn{)}

\mbox{\hphantom{\Scribtexttt{xxxx}}}\RktPn{(}\RktSym{define/public}\mbox{\hphantom{\Scribtexttt{x}}}\RktPn{(}\RktSym{dependencies}\RktPn{)}\mbox{\hphantom{\Scribtexttt{x}}}\RktPn{(}\RktSym{list}\mbox{\hphantom{\Scribtexttt{x}}}\RktSym{@label}\RktPn{)}\RktPn{)}

\mbox{\hphantom{\Scribtexttt{xxxx}}}\RktPn{(}\RktSym{define/public}\mbox{\hphantom{\Scribtexttt{x}}}\RktPn{(}\RktSym{create}\mbox{\hphantom{\Scribtexttt{x}}}\RktSym{parent}\RktPn{)}

\mbox{\hphantom{\Scribtexttt{xxxxxx}}}\RktPn{(}\RktSym{new}\mbox{\hphantom{\Scribtexttt{x}}}\RktSym{gui{\hbox{\texttt{:}}}message\%}\mbox{\hphantom{\Scribtexttt{x}}}\RktPn{[}\RktSym{parent}\mbox{\hphantom{\Scribtexttt{x}}}\RktSym{parent}\RktPn{]}

\mbox{\hphantom{\Scribtexttt{xxxxxxxxxxx}}}\RktPn{[}\RktSym{label}\mbox{\hphantom{\Scribtexttt{x}}}\RktPn{(}\RktSym{obs{-}peek}\mbox{\hphantom{\Scribtexttt{x}}}\RktSym{@label}\RktPn{)}\RktPn{]}\RktPn{)}\RktPn{)}

\mbox{\hphantom{\Scribtexttt{xxxx}}}\RktPn{(}\RktSym{define/public}\mbox{\hphantom{\Scribtexttt{x}}}\RktPn{(}\RktSym{update}\mbox{\hphantom{\Scribtexttt{x}}}\RktSym{widget}\mbox{\hphantom{\Scribtexttt{x}}}\RktSym{what}\mbox{\hphantom{\Scribtexttt{x}}}\RktSym{val}\RktPn{)}

\mbox{\hphantom{\Scribtexttt{xxxxxx}}}\RktPn{(}\RktSym{send}\mbox{\hphantom{\Scribtexttt{x}}}\RktSym{widget}\mbox{\hphantom{\Scribtexttt{x}}}\RktSym{set{-}label}\mbox{\hphantom{\Scribtexttt{x}}}\RktSym{val}\RktPn{)}\RktPn{)}

\mbox{\hphantom{\Scribtexttt{xxxx}}}\RktPn{(}\RktSym{define/public}\mbox{\hphantom{\Scribtexttt{x}}}\RktPn{(}\RktSym{destroy}\mbox{\hphantom{\Scribtexttt{x}}}\RktSym{widget}\RktPn{)}\mbox{\hphantom{\Scribtexttt{x}}}\RktPn{(}\RktSym{void}\RktPn{)}\RktPn{)}\RktPn{)}\RktPn{)}

\mbox{\hphantom{\Scribtexttt{x}}}

\RktPn{(}\RktSym{define}\mbox{\hphantom{\Scribtexttt{x}}}\RktPn{(}\RktSym{text}\mbox{\hphantom{\Scribtexttt{x}}}\RktSym{@label}\RktPn{)}

\mbox{\hphantom{\Scribtexttt{xx}}}\RktPn{(}\RktSym{new}\mbox{\hphantom{\Scribtexttt{x}}}\RktSym{text\%}\mbox{\hphantom{\Scribtexttt{x}}}\RktPn{[}\RktSym{@label}\mbox{\hphantom{\Scribtexttt{x}}}\RktSym{@label}\RktPn{]}\RktPn{)}\RktPn{)}\end{SingleColumn}\end{RktBlk}\end{SCodeFlow}\end{FigureInside}\end{Centerfigure}

\Centertext{\Legend{\FigureTarget{\label{t:x28counter_x28x22figurex22_x22viewx2dimplx2erktx22x29x29}\textbf{Figure}~\textbf{7}. }{t:x28counter_x28x22figurex22_x22viewx2dimplx2erktx22x29x29}An implementation of a custom \RktSym{view{\Stttextless}\%{\Stttextmore}} for displaying label text.}}\end{Figure}

To go from a \RktSym{view{\Stttextless}\%{\Stttextmore}} to a functional view, all that remains
is to wrap object construction in a function. Thus, the shell{---}the
part that most library consumers interact with{---}is functional.
Figure~\hyperref[t:x28counter_x28x22figurex22_x22viewx2dimplx2erktx22x29x29]{\FigureRef{7}{t:x28counter_x28x22figurex22_x22viewx2dimplx2erktx22x29x29}} shows an implementation of a custom
\RktSym{view{\Stttextless}\%{\Stttextmore}} and its function wrapper.

How does such a shell permit the use of functional programming
techniques? We have already seen in the previous sections and in code
examples that this shell abstracts away all the imperative details from
most library consumers: until now, we have not needed to understand
the imperative object{-}based API being wrapped in order to write GUI
programs. Further, those GUI programs have used functional programming
techniques, such as composition of reusable views. Even the Frosthaven
Manager sticks mostly to GUI Easy{'}s functional shell and is thus able to
use the {``}Functional Core, Imperative Shell{''} architecture.

The key lesson for functional programmers here is that, when possible,
wrapping an imperative API in a functional shell enables all the
benefits of functional programming. For highly complex systems, like
GUIs, to rewrite the entire system in a functional style may be
impractical. Instead, it is more practical to reuse existing imperative
or object{-}based work by wrapping it in a functional shell.

\Ssubsection{Inversion of Control}{Inversion of Control}\label{t:x28part_x22iocx22x29}

Inversion of control refers to an architecture wherein the main
application provides procedures called by some framework, rather than by
other application code. The framework is responsible for most of the
coordinating activity, such as managing an event loop\Autobibref{~[\hyperref[t:x28autobib_x22Ralph_Ex2e_Johnson_and_Brian_FooteDesigning_Reusable_ClassesJournal_of_Objectx2dOriented_Programming_1x282x29x2c_ppx2e_22x2dx2d351988httpx3ax2fx2fwwwx2elaputanx2eorgx2fdrcx2fdrcx2ehtmlx22x29]{\AutobibLink{19}}]}.

GUI Easy is one such framework. It manages the object lifecyle of
\RktSym{view{\Stttextless}\%{\Stttextmore}} instances, which is also the lifecyle of GUI Easy
graphical applications. Calling \RktSym{render} on a view kicks off that
lifecycle, which is managed by the GUI Easy library and which calls into
application code so that it may respond to user interaction.

Inversion of control leads to an extensible application skeleton: the
backbone of the application is under the framework{'}s control.
Applications hang the meat of their tasks on the extension points
provided by the framework. In the case of GUI Easy, those extension
points are (a) the event handlers provided by standard components, also
called the controllers in \SecRef{\SectionNumberLink{t:x28part_x22mvcx22x29}{3.3}}{Models, Views, and Controllers}, and (b) the \RktSym{view{\Stttextless}\%{\Stttextmore}}
interface for creating new framework{-}aware components. Being able to
compose individual framework components into larger components also
contributes to extensibility and reuse.

\Ssubsection{Challenges}{Challenges}\label{t:x28part_x22Challengesx22x29}

Naturally, maintaining reusable components and programming against a
functional shell is not without its challenges. What do you do when you
need access to the underlying object{-}oriented API for a feature not
exposed by existing wrappers? How do you handle a piece of nearly{-}global
state whose usage is hard to predict when writing reusable components?
Fortunately, both of these problems have solutions.

\begin{Figure}\begin{Centerfigure}\begin{FigureInside}\begin{RktBlk}\begin{SingleColumn}\RktModLink{\RktMod{\#lang}}\mbox{\hphantom{\Scribtexttt{x}}}\RktModLink{\RktSym{racket/gui/easy}}

\RktPn{(}\RktSym{require}\mbox{\hphantom{\Scribtexttt{x}}}\RktSym{racket/class}\RktPn{)}

\RktPn{(}\RktSym{define}\mbox{\hphantom{\Scribtexttt{x}}}\RktSym{close{\hbox{\texttt{!}}}}\mbox{\hphantom{\Scribtexttt{x}}}\RktSym{void}\RktPn{)}

\RktPn{(}\RktSym{render}

\mbox{\hphantom{\Scribtexttt{x}}}\RktPn{(}\RktSym{window}

\mbox{\hphantom{\Scribtexttt{xx}}}\RktPn{\#{\hbox{\texttt{:}}}title}\mbox{\hphantom{\Scribtexttt{x}}}\RktVal{"Goodbye World"}

\mbox{\hphantom{\Scribtexttt{xx}}}\RktPn{\#{\hbox{\texttt{:}}}mixin}\mbox{\hphantom{\Scribtexttt{x}}}\RktPn{(}\RktSym{$\lambda$}\mbox{\hphantom{\Scribtexttt{x}}}\RktPn{(}\RktSym{window\%}\RktPn{)}

\mbox{\hphantom{\Scribtexttt{xxxxxxxxxxxx}}}\RktPn{(}\RktSym{class}\mbox{\hphantom{\Scribtexttt{x}}}\RktSym{window\%}\mbox{\hphantom{\Scribtexttt{x}}}\RktPn{(}\RktSym{super{-}new}\RktPn{)}

\mbox{\hphantom{\Scribtexttt{xxxxxxxxxxxxxx}}}\RktPn{(}\RktSym{set{\hbox{\texttt{!}}}}\mbox{\hphantom{\Scribtexttt{x}}}\RktSym{close{\hbox{\texttt{!}}}}

\mbox{\hphantom{\Scribtexttt{xxxxxxxxxxxxxxxx}}}\RktPn{(}\RktSym{$\lambda$}\mbox{\hphantom{\Scribtexttt{x}}}\RktPn{(}\RktPn{)}

\mbox{\hphantom{\Scribtexttt{xxxxxxxxxxxxxxxxxx}}}\RktPn{(}\RktSym{when}\mbox{\hphantom{\Scribtexttt{x}}}\RktPn{(}\RktSym{send}\mbox{\hphantom{\Scribtexttt{x}}}\RktSym{this}\mbox{\hphantom{\Scribtexttt{x}}}\RktSym{can{-}close{\hbox{\texttt{?}}}}\RktPn{)}

\mbox{\hphantom{\Scribtexttt{xxxxxxxxxxxxxxxxxxxx}}}\RktPn{(}\RktSym{send}\mbox{\hphantom{\Scribtexttt{x}}}\RktSym{this}\mbox{\hphantom{\Scribtexttt{x}}}\RktSym{on{-}close}\RktPn{)}

\mbox{\hphantom{\Scribtexttt{xxxxxxxxxxxxxxxxxxxx}}}\RktPn{(}\RktSym{send}\mbox{\hphantom{\Scribtexttt{x}}}\RktSym{this}\mbox{\hphantom{\Scribtexttt{x}}}\RktSym{show}\mbox{\hphantom{\Scribtexttt{x}}}\RktVal{\#f}\RktPn{)}\RktPn{)}\RktPn{)}\RktPn{)}\RktPn{)}\RktPn{)}

\mbox{\hphantom{\Scribtexttt{xx}}}\RktPn{(}\RktSym{button}\mbox{\hphantom{\Scribtexttt{x}}}\RktVal{"Click Me{\hbox{\texttt{!}}}"}\mbox{\hphantom{\Scribtexttt{x}}}\RktPn{(}\RktSym{$\lambda$}\mbox{\hphantom{\Scribtexttt{x}}}\RktPn{(}\RktPn{)}\mbox{\hphantom{\Scribtexttt{x}}}\RktPn{(}\RktSym{close{\hbox{\texttt{!}}}}\RktPn{)}\RktPn{)}\RktPn{)}\RktPn{)}\RktPn{)}\end{SingleColumn}\end{RktBlk}\end{FigureInside}\end{Centerfigure}

\Centertext{\Legend{\FigureTarget{\label{t:x28counter_x28x22figurex22_x22goodbyex2dworldx2erktx22x29x29}\textbf{Figure}~\textbf{8}. }{t:x28counter_x28x22figurex22_x22goodbyex2dworldx2erktx22x29x29}Using mixins to write a GUI Easy app whose window is closed when a button is clicked.}}\end{Figure}

The problem of access to imperative behaviors is solved by GUI Easy
conventions. In an object{-}oriented toolkit, we would subclass widgets as
needed to create new behaviors. We cannot subclass a class we cannot
access, for it is ostensibly hidden by the wrapper. In response, some
GUI Easy views support a mixin\Autobibref{~[\hyperref[t:x28autobib_x22Gilad_BrachaThe_Programming_Languages_Jigsawx3a_Mixinsx2c_Modularityx2c_and_InheritancePhD_dissertationx2c_University_of_Utah1992httpsx3ax2fx2fbrachax2eorgx2fjigsawx2epdfChx2e_3x22x29]{\AutobibLink{3}}, \hyperref[t:x28autobib_x22William_Rx2e_CookA_Denotational_Semantics_of_InheritancePhD_dissertationx2c_Brown_University1989httpsx3ax2fx2fwwwx2ecsx2eutexasx2eedux2fx7ewcookx2fpapersx2fthesisx2fcook89x2epdfChx2e_10x22x29]{\AutobibLink{5}}, \hyperref[t:x28autobib_x22Matthew_Flattx2c_Shriram_Krishnamurthix2c_and_Matthias_FelleisenClasses_and_mixinsIn_Procx2e_25th_ACM_SIGPLANx2dSIGACT_symposium_on_Principles_of_programming_languagesx2c_POPL_x2798x2c_ppx2e_171x2dx2d1831998doix3ahttpsx3ax2fx2fdoix2eorgx2f10x2e1145x2f268946x2e268961x22x29]{\AutobibLink{13}}, \hyperref[t:x28autobib_x22David_Goldbergx2c_Robert_Bruce_Findlerx2c_and_Matthew_FlattSuper_and_Innerx2dx2dx2dTogether_at_Lastx21In_Procx2e_Objectx2dOriented_Programmingx2c_Languagesx2c_Systemsx2c_and_Applications2004httpx3ax2fx2fwwwx2ecsx2eutahx2eedux2fpltx2fpublicationsx2foopsla04x2dgffx2epdfx22x29]{\AutobibLink{16}}, \hyperref[t:x28autobib_x22David_Ax2e_MoonObjectx2doriented_programming_with_FlavorsIn_Procx2e_ACM_Conference_on_Objectx2doriented_Programmingx2c_Systemsx2c_Languagesx2c_and_Applicationsx2c_ppx2e_1x2dx2d81986httpsx3ax2fx2fwwwx2ecsx2etuftsx2eedux2fcompx2f150FPx2farchivex2fdavidx2dmoonx2fflavorsx2epdfx22x29]{\AutobibLink{23}}]} argument, a function from class
to class. Mixins allows us to dynamically subclass widgets at runtime to
override or augment their methods. \hyperref[t:x28autobib_x22Brad_Ax2e_MyersA_New_Model_for_Handling_InputACM_Transactions_on_Information_Systems_8x283x29x2c_ppx2e_289x2dx2d3201990doix3ahttpsx3ax2fx2fdoix2eorgx2f10x2e1145x2f98188x2e98204x22x29]{\AutobibLink{Myers}}{'} {``}Goodbye
World{''} program provides a good example: how can we include a button in
the view that closes the window when such functionality is only present
in the object{-}oriented toolkit? Figure~\hyperref[t:x28counter_x28x22figurex22_x22goodbyex2dworldx2erktx22x29x29]{\FigureRef{8}{t:x28counter_x28x22figurex22_x22goodbyex2dworldx2erktx22x29x29}} shows
how: by using a mixin, we can get a reference to the window{'}s
\RktSym{on{-}close} and \RktSym{show} methods. The Frosthaven Manager uses
mixins to implement window close behavior like in the {``}Goodbye World{''}
program combined with a macro that implements the
mixin{-}over{-}\RktPn{(}\RktSym{set{\hbox{\texttt{!}}}}\Scribtexttt{ }\RktSym{close{\hbox{\texttt{!}}}}\Scribtexttt{ }\RktSym{{\hbox{\texttt{.}}}{\hbox{\texttt{.}}}{\hbox{\texttt{.}}}}\RktPn{)} pattern; it also augments window
close behavior so that closing the window can behave like accepting a
choice. When mixins are insufficient, we can choose to write our own
\RktSym{view{\Stttextless}\%{\Stttextmore}} implementation to wrap any widget we desire. The
Frosthaven Manager uses custom \RktSym{view{\Stttextless}\%{\Stttextmore}}s to display rendered
Markdown\Autobibref{~[\hyperref[t:x28autobib_x22John_GruberDaring_Fireballx3a_Markdown2023httpsx3ax2fx2fdaringfireballx2enetx2fprojectsx2fmarkdownx2fRetrieved_June_2023x2ex22x29]{\AutobibLink{17}}]} files, for example.

\begin{Figure}\begin{Centerfigure}\begin{FigureInside}\begin{SCodeFlow}\begin{RktBlk}\begin{SingleColumn}\RktPn{(}\RktSym{define}\mbox{\hphantom{\Scribtexttt{x}}}\RktPn{(}\RktSym{monster{-}group{-}view}\mbox{\hphantom{\Scribtexttt{x}}}\RktSym{@monsters}\mbox{\hphantom{\Scribtexttt{x}}}\RktSym{@env}\RktPn{)}

\mbox{\hphantom{\Scribtexttt{xx}}}\RktPn{(}\RktSym{define}\mbox{\hphantom{\Scribtexttt{x}}}\RktSym{@monster}\mbox{\hphantom{\Scribtexttt{x}}}\RktSym{{\hbox{\texttt{.}}}{\hbox{\texttt{.}}}{\hbox{\texttt{.}}}}\RktPn{)}

\mbox{\hphantom{\Scribtexttt{xx}}}\RktPn{(}\RktSym{tabs}\mbox{\hphantom{\Scribtexttt{x}}}\RktSym{@monsters}

\mbox{\hphantom{\Scribtexttt{xxxxxxxx}}}\RktPn{(}\RktSym{monster{-}view}\mbox{\hphantom{\Scribtexttt{x}}}\RktSym{@monster}\mbox{\hphantom{\Scribtexttt{x}}}\RktSym{@env}\RktPn{)}\RktPn{)}\RktPn{)}

\mbox{\hphantom{\Scribtexttt{x}}}

\RktPn{(}\RktSym{define}\mbox{\hphantom{\Scribtexttt{x}}}\RktPn{(}\RktSym{monster{-}view}\mbox{\hphantom{\Scribtexttt{x}}}\RktSym{@monster}\mbox{\hphantom{\Scribtexttt{x}}}\RktSym{@env}\RktPn{)}

\mbox{\hphantom{\Scribtexttt{xx}}}\RktPn{(}\RktSym{counter}\mbox{\hphantom{\Scribtexttt{x}}}\RktPn{(}\RktSym{monster{-}{\Stttextmore}hp{-}text}\mbox{\hphantom{\Scribtexttt{x}}}\RktSym{@monster}\mbox{\hphantom{\Scribtexttt{x}}}\RktSym{@env}\RktPn{)}

\mbox{\hphantom{\Scribtexttt{xxxxxxxxxxx}}}\RktPn{(}\RktSym{$\lambda$}\mbox{\hphantom{\Scribtexttt{x}}}\RktPn{(}\RktSym{action}\RktPn{)}\mbox{\hphantom{\Scribtexttt{x}}}\RktSym{{\hbox{\texttt{.}}}{\hbox{\texttt{.}}}{\hbox{\texttt{.}}}}\RktPn{)}\RktPn{)}\RktPn{)}\end{SingleColumn}\end{RktBlk}\end{SCodeFlow}\end{FigureInside}\end{Centerfigure}

\Centertext{\Legend{\FigureTarget{\label{t:x28counter_x28x22figurex22_x22threadingx2erktx22x29x29}\textbf{Figure}~\textbf{9}. }{t:x28counter_x28x22figurex22_x22threadingx2erktx22x29x29}Threading the \RktSym{@env} argument from a view for monster groups to a view for a monster to a view for a monster{\textquotesingle}s hit points.}}\end{Figure}

The problem of global state is handled by functional programming
techniques. Essentially, we have two choices: threading state or
dynamic binding. If we are confident that the state will be required
in all reusable views, we can thread the state as input from one view
to the next, like threading a needle through all parts of the program.
Threaded state is the solution preferred by DDAU and reusuable views.
For example, the Frosthaven Manager threads an observable
\RktSym{@env} throughout the application so that simple arithmetic
formulas with variables can be evaluated for monster information or
scenario{-}specific attributes. As a result, many views take a
\RktSym{@env} argument, and many views pass a \RktSym{@env} to child
views. Figure~\hyperref[t:x28counter_x28x22figurex22_x22threadingx2erktx22x29x29]{\FigureRef{9}{t:x28counter_x28x22figurex22_x22threadingx2erktx22x29x29}} shows a simplified example.

Threading rarely{-}used state quickly becomes tedious and, when not needed
everywhere, tangles unnecessary concerns. In response, we can use
dynamic binding, which breaks some functional purity for convenience
and allows us to refer to external state. Using dynamic binding makes
views less reusable: they now have dependencies not defined by their
inputs. Dynamic binding permits each view to only be concerned with
the global state if absolutely necessary. The Frosthaven Manager
threads state as much as possible but does use dynamic binding in rare
instances. It is important to mention that using dynamic binding via
Racket{'}s parameters is not straightforward when working with the GUI
system due to the multi{-}threaded environment and queued callbacks; to
achieve dynamic{-}binding for the Frosthaven Manager, Ben had to both bind
parameters in the GUI event threads and take care to spawn more event
threads when new bindings were needed. This complexity may not be worth
it in all applications.

\sectionNewpage

\Ssection{Related Work}{Related Work}\label{t:x28part_x22relatedx5fworkx22x29}

GUI Easy draws inspiration from Swift UI\Autobibref{~[\hyperref[t:x28autobib_x22AppleSwiftUI2023httpsx3ax2fx2fdeveloperx2eapplex2ecomx2fxcodex2fswiftuix2fRetrieved_June_2023x2ex22x29]{\AutobibLink{1}}]}, another
system that wraps an imperative GUI framework in a functional shell.
Other sources of inspiration include Clojure{'}s Reagent\Autobibref{~[\hyperref[t:x28autobib_x22reagentx2dprojectReagent2023httpsx3ax2fx2fgithubx2ecomx2freagentx2dprojectx2freagentRetrieved_June_2023x2ex22x29]{\AutobibLink{28}}]}
and JavaScript{'}s React\Autobibref{~[\hyperref[t:x28autobib_x22Meta_Open_SourceReact2023httpsx3ax2fx2freactx2edevRetrieved_June_2023x2ex22x29]{\AutobibLink{29}}]}.

The Elm\Autobibref{~[\hyperref[t:x28autobib_x22Evan_CzaplickiElm_x2d_delightful_language_for_reliable_web_applications2021httpsx3ax2fx2felmx2dlangx2eorgRetrieved_June_2023x2ex22x29]{\AutobibLink{10}}]} programming language strictly constrains
component composition to the data down, actions up style. Clojure{'}s
re{-}frame\Autobibref{~[\hyperref[t:x28autobib_x22Day_8_Technologyrex2dframe2023httpsx3ax2fx2fgithubx2ecomx2fday8x2frex2dframeRetrieved_June_2023x2ex22x29]{\AutobibLink{31}}]} library builds on Reagent\Autobibref{~[\hyperref[t:x28autobib_x22reagentx2dprojectReagent2023httpsx3ax2fx2fgithubx2ecomx2freagentx2dprojectx2freagentRetrieved_June_2023x2ex22x29]{\AutobibLink{28}}]}
to add more sophisticated state management. This includes a global
store and effect handler, akin to GUI Easy{'}s observables and update
procedures, and queries, akin to GUI Easy{'}s derived observables.

Frapp\'{e}\Autobibref{~[\hyperref[t:x28autobib_x22Antony_CourtneyFrappxe9x3a_Functional_Reactive_Programming_in_JavaIn_Procx2e_Practical_Aspects_of_Declarative_Languages2001httpsx3ax2fx2fdoix2eorgx2f10x2e1007x2f3x2d540x2d45241x2d9x5f3x22x29]{\AutobibLink{9}}]} is an implementation of FRP in Java that wraps an
imperative API (Java Beans) in a declarative shell. Both GUI Easy and
Frapp\'{e} implement a {``}push{''} model for propagation of values through the
dependency graph: \textit{behaviors} in Frapp\'{e} hold values and support
the registration of listeners to be notified when their held values
change, like observables in GUI Easy. Unlike Frapp\'{e}, GUI Easy does not
have an explicit notion of \textit{events}. Instead, observables may be
directly updated in response to callbacks.

In Racket, FrTime\Autobibref{~[\hyperref[t:x28autobib_x22Gregory_Cooper_and_Shriram_KrishnamurthiFrTimex3a_Functional_Reactive_Programming_in_PLT_SchemeBrownx2c_CSx2d03x2d202004httpsx3ax2fx2fcsx2ebrownx2eedux2fresearchx2fpubsx2ftechreportsx2freportsx2fCSx2d03x2d20x2ehtmlx22x29]{\AutobibLink{6}}]} implements a push{-}based
FRP language for GUIs and other tasks. The FrTime language extends a
subset of the Racket language to make signal values first{-}class. By
contrast, GUI Easy is a regular library built on top of the Racket
language{---}a conscious choice in order to make it straightforward to
bring GUI Easy into existing Racket programs. One symptom of this choice
is that, while FrTime signals can be displayed in a continuously
variable manner with support from editors like DrRacket, GUI Easy
observables are regular Racket values and are displayed as such. FrTime
and GUI Easy both track state by using mutation internally and both
FrTime \textit{behaviors} and GUI Easy observables get updated
asynchronously in response to changes.

Fred\Autobibref{~[\hyperref[t:x28autobib_x22Daniel_Ignatoffx2c_Gregory_Hx2e_Cooperx2c_and_Shriram_KrishnamurthiCrossing_State_Linesx3a_Adapting_Objectx2dOriented_Frameworks_to_Functional_Reactive_LanguagesIn_Procx2e_Functional_and_Logic_Programmingx2c_FLOPS_20062006httpsx3ax2fx2flinkx2espringerx2ecomx2fchapterx2f10x2e1007x2f11737414x5f18x22x29]{\AutobibLink{18}}]} is FrTime{'}s wrapper around Racket GUI. It
wraps the object{-}oriented API of Racket GUI by subclassing Racket GUI
widgets to work with FrTime signal values. By contrast, GUI Easy views
are separate classes that implement the \RktSym{view{\Stttextless}\%{\Stttextmore}} interface.
Despite this difference, both frameworks perform similar operations
in order to connect their reactive abstractions to the underlying
widgets. FrTime makes use of macros to generate most of its wrapper
code, whereas GUI Easy views are implemented manually. Unlike GUI Easy,
Fred does not hide the details of the Racket class system from the end
user. Because its widgets sublass Racket GUI widgets, it has the same
order{-}of{-}definition constraints as Racket GUI that we described in
\SecRef{\SectionNumberLink{t:x28part_x22questx2dforx2dguix2deasyx22x29}{2.1}}{Quest for Easier GUIs}.

Flapjax\Autobibref{~[\hyperref[t:x28autobib_x22Leo_Ax2e_Meyerovichx2c_Arjun_Guhax2c_Jacob_Baskinx2c_Gregory_Hx2e_Cooperx2c_Michael_Greenbergx2c_Aleks_Bromfieldx2c_and_Shriram_KrishnamurthiFlapjaxx3a_a_programming_language_for_Ajax_applicationsIn_Procx2e_ACM_SIGPLAN_conference_on_Object_oriented_programming_systems_languages_and_applicationsx2c_OOPSLA_20092009httpsx3ax2fx2fdlx2eacmx2eorgx2fdoix2f10x2e1145x2f1640089x2e1640091x22x29]{\AutobibLink{22}}]} is a push{-}based FRP implementation. It provides
both a compiler from the Flapjax language to JavaScript and a standalone
library. Similar to FrTime for Racket, the Flapjax language extends
JavaScript to make behaviors first{-}class, implicitly lifting expressions
to work over behaviors where necessary. The Flapjax compiler is optional
and the standalone library can be used directly from JavaScript without
compiler support, like GUI Easy can be from Racket. While GUI Easy
observables get updated asynchronously and independently, updates in
Flapjax are propagated through the dependency graph in topological
order, avoiding potential inconsistencies between behaviors that share
part of the dependency graph.

The Andrew toolkit and Garnet system, among others of that time, knew
that the MVC architecture tightly couples views and
controllers\Autobibref{~[\hyperref[t:x28autobib_x22Brad_Ax2e_MyersA_New_Model_for_Handling_InputACM_Transactions_on_Information_Systems_8x283x29x2c_ppx2e_289x2dx2d3201990doix3ahttpsx3ax2fx2fdoix2eorgx2f10x2e1145x2f98188x2e98204x22x29]{\AutobibLink{24}}, \hyperref[t:x28autobib_x22Andrew_Jx2e_Palayx2c_Wilfred_Jx2e_Hansenx2c_Michael_Lx2e_Kazarx2c_Mark_Shermanx2c_Maria_Gx2e_Wadlowx2c_Thomas_Px2e_Neuendorfferx2c_Zalman_Sternx2c_Miles_Baderx2c_and_Thom_PetersThe_Andrew_Toolkitx2014An_OverviewIn_Procx2e_USENIX_Winter_Conferencex2c_ppx2e_9x2dx2d221988x22x29]{\AutobibLink{25}}]}. Typical solutions involve not
separating views and controllers or dropping the controller
altogether\Autobibref{~[\hyperref[t:x28autobib_x22Brad_Ax2e_MyersA_New_Model_for_Handling_InputACM_Transactions_on_Information_Systems_8x283x29x2c_ppx2e_289x2dx2d3201990doix3ahttpsx3ax2fx2fdoix2eorgx2f10x2e1145x2f98188x2e98204x22x29]{\AutobibLink{24}}]}. DDAU from \SecRef{\SectionNumberLink{t:x28part_x22Reusablex5fViewsx22x29}{5.1}}{Reusable Views} encourages
decoupling the view and controller by the use of callbacks: they provide
the same interposition points a typical controller would use to respond
to user interaction, and they provide different view instances the
ability to respond with different model updates. This is especially
important when the view can display many different models. Decoupling
views and controllers also allow combining controllers when combining
views. In the Garnet system, however, {``}spaghetti{''} callbacks are
avoided by providing a small set of flexible Interactors and by using
formulated constraints to tie together interactions and
updates\Autobibref{~[\hyperref[t:x28autobib_x22Brad_Ax2e_MyersA_New_Model_for_Handling_InputACM_Transactions_on_Information_Systems_8x283x29x2c_ppx2e_289x2dx2d3201990doix3ahttpsx3ax2fx2fdoix2eorgx2f10x2e1145x2f98188x2e98204x22x29]{\AutobibLink{24}}]}.

Inversion of control has a long history: the development of the
Tajo\Autobibref{~[\hyperref[t:x28autobib_x22Donald_Cx2e_WallaceTajo_Functional_Specification_Version_6x2e0x2cXerox_Internal_Document1980x22x29]{\AutobibLink{32}}]} and Mesa\Autobibref{~[\hyperref[t:x28autobib_x22Richard_Ex2e_SweetThe_Mesa_Programming_EnvironmentACM_SIGPLAN_Notices_20x287x29x2c_ppx2e_216x2dx2d2291985doix3ahttpsx3ax2fx2fdoix2eorgx2f10x2e1145x2f17919x2e806843x22x29]{\AutobibLink{30}}]} systems called it the
{``}Hollywood Principle.{''} \hyperref[t:x28autobib_x22Brad_Ax2e_MyersA_New_Model_for_Handling_InputACM_Transactions_on_Information_Systems_8x283x29x2c_ppx2e_289x2dx2d3201990doix3ahttpsx3ax2fx2fdoix2eorgx2f10x2e1145x2f98188x2e98204x22x29]{\AutobibLink{Myers}}, developing the Garnet
system, similarly separated monitoring processes from the users
application code, providing hooks for the application to respond to
events from the framework\Autobibref{~[\hyperref[t:x28autobib_x22Brad_Ax2e_MyersA_New_Model_for_Handling_InputACM_Transactions_on_Information_Systems_8x283x29x2c_ppx2e_289x2dx2d3201990doix3ahttpsx3ax2fx2fdoix2eorgx2f10x2e1145x2f98188x2e98204x22x29]{\AutobibLink{24}}]}.

In the language of \Autobibref{\hyperref[t:x28autobib_x22Ralph_Ex2e_Johnson_and_Brian_FooteDesigning_Reusable_ClassesJournal_of_Objectx2dOriented_Programming_1x282x29x2c_ppx2e_22x2dx2d351988httpx3ax2fx2fwwwx2elaputanx2eorgx2fdrcx2fdrcx2ehtmlx22x29]{\AutobibLink{Johnson and Foote}}~[\hyperref[t:x28autobib_x22Ralph_Ex2e_Johnson_and_Brian_FooteDesigning_Reusable_ClassesJournal_of_Objectx2dOriented_Programming_1x282x29x2c_ppx2e_22x2dx2d351988httpx3ax2fx2fwwwx2elaputanx2eorgx2fdrcx2fdrcx2ehtmlx22x29]{\AutobibLink{19}}]}, we ask whether GUI Easy is a
{``}white{-}box{''} or {``}black{-}box{''} inversion{-}of{-}control framework. White{-}box
frameworks, so{-}called because they are transparent, typically require
programs to subclass and add methods to framework components, which
requires understanding implementation details. In contrast, black{-}box
frameworks are an {``}evolutionary goal,{''} in which opaque components
communicate only via a shared protocol. Black{-}box frameworks are thus
easier to learn and use. White{-}box frameworks typically maintain global
state, while black{-}box frameworks see state shared explicitly when
needed. Given these criteria, we can confidently state that the
object{-}oriented Racket GUI toolkit is a white{-}box framework. GUI Easy is
principally black{-}box, relying on a protocol of observables for state
and procedures for communication. Yet GUI Easy provides escape hatches
of varying complexity that bring back the expert flavor of white{-}box
frameworks; namely, mixins and the \RktSym{view{\Stttextless}\%{\Stttextmore}} interface. To give
credit where it is due, we recognize that abstracting is much easier
given a plethora of worked examples{---}we would not have the experience
to develop GUI Easy without Racket{'}s GUI toolkit. On the flipside, using
GUI Easy has proven to be a good way to learn how to use Racket{'}s GUI
toolkit!

\sectionNewpage

\Ssection{Conclusion}{Conclusion}\label{t:x28part_x22Conclusionx22x29}

We have reported on the difficulties of programming stateful GUIs
with imperative, object{-}based APIs. We also described a functional
wrapper around Racket{'}s object{-}oriented GUI library that aims to solve
some of those shortcomings. GUI Easy has been successfully used for
small and large projects, including the Frosthaven Manager discussed
in this report. We derived several architectural principles from the
construction of both projects: functional shells over imperative APIs
enable programmers to use functional programming techniques even when
dealing with a system whose underlying implementation is imperative.
Extensible hooks are necessary in functional shells to permit access
to the underlying systems where needed. Reusable components, much like
pure functions, should not mutate external state. Like pure functions,
reusable components are independently testable and are easily composed
with one another.

\begin{acks}

Ben is grateful to Savannah Knoble, Derrick Franklin, John Hines,
and Jake Hicks for playtesting the Frosthaven Manager throughout
development, and to Isaac Childres for bringing us the wonderful world
of Frosthaven.

We thank the anonymous reviewers, our shepherd Shriram Krishnamurthi,
and our early readers Jeff Terrell, Marc Kaufmann, Matthew Flatt, and
Robby Findler for their insightful comments.

\end{acks}

\sectionNewpage

\Ssectionstarx{References}{References}\label{t:x28part_x22docx2dbibliographyx22x29}

\begin{bigtabular}{@{\bigtableleftpad}l@{}l@{}}
\hbox{\Autocolbibnumber{[1]}} &
\hbox{\Autobibtarget{\label{t:x28autobib_x22AppleSwiftUI2023httpsx3ax2fx2fdeveloperx2eapplex2ecomx2fxcodex2fswiftuix2fRetrieved_June_2023x2ex22x29}\Autocolbibentry{Apple. SwiftUI. 2023. \href{https://developer.apple.com/xcode/swiftui/}{\Snolinkurl{https://developer.apple.com/xcode/swiftui/}} Retrieved June 2023.}}} \\
\hbox{\Autocolbibnumber{[2]}} &
\hbox{\Autobibtarget{\label{t:x28autobib_x22Gary_BernhardtFunctional_Corex2c_Imperative_Shell2012httpsx3ax2fx2fwwwx2edestroyallsoftwarex2ecomx2fscreencastsx2fcatalogx2ffunctionalx2dcorex2dimperativex2dshellx22x29}\Autocolbibentry{Gary Bernhardt. Functional Core, Imperative Shell. 2012. \href{https://www.destroyallsoftware.com/screencasts/catalog/functional-core-imperative-shell}{\Snolinkurl{https://www.destroyallsoftware.com/screencasts/catalog/functional-core-imperative-shell}}}}} \\
\hbox{\Autocolbibnumber{[3]}} &
\hbox{\Autobibtarget{\label{t:x28autobib_x22Gilad_BrachaThe_Programming_Languages_Jigsawx3a_Mixinsx2c_Modularityx2c_and_InheritancePhD_dissertationx2c_University_of_Utah1992httpsx3ax2fx2fbrachax2eorgx2fjigsawx2epdfChx2e_3x22x29}\Autocolbibentry{Gilad Bracha. The Programming Languages Jigsaw: Mixins, Modularity, and Inheritance. PhD dissertation, University of Utah, 1992. \href{https://bracha.org/jigsaw.pdf}{\Snolinkurl{https://bracha.org/jigsaw.pdf}} Ch. 3}}} \\
\hbox{\Autocolbibnumber{[4]}} &
\hbox{\Autobibtarget{\label{t:x28autobib_x22Cephalofair_GamesFrosthaven2023httpsx3ax2fx2fcephalofairx2ecomx2fpagesx2ffrosthavenx22x29}\Autocolbibentry{Cephalofair Games. Frosthaven. 2023. \href{https://cephalofair.com/pages/frosthaven}{\Snolinkurl{https://cephalofair.com/pages/frosthaven}}}}} \\
\hbox{\Autocolbibnumber{[5]}} &
\hbox{\Autobibtarget{\label{t:x28autobib_x22William_Rx2e_CookA_Denotational_Semantics_of_InheritancePhD_dissertationx2c_Brown_University1989httpsx3ax2fx2fwwwx2ecsx2eutexasx2eedux2fx7ewcookx2fpapersx2fthesisx2fcook89x2epdfChx2e_10x22x29}\Autocolbibentry{William R. Cook. A Denotational Semantics of Inheritance. PhD dissertation, Brown University, 1989. \href{https://www.cs.utexas.edu/~wcook/papers/thesis/cook89.pdf}{\Snolinkurl{https://www.cs.utexas.edu/~wcook/papers/thesis/cook89.pdf}} Ch. 10}}} \\
\hbox{\Autocolbibnumber{[6]}} &
\hbox{\Autobibtarget{\label{t:x28autobib_x22Gregory_Cooper_and_Shriram_KrishnamurthiFrTimex3a_Functional_Reactive_Programming_in_PLT_SchemeBrownx2c_CSx2d03x2d202004httpsx3ax2fx2fcsx2ebrownx2eedux2fresearchx2fpubsx2ftechreportsx2freportsx2fCSx2d03x2d20x2ehtmlx22x29}\Autocolbibentry{Gregory Cooper and Shriram Krishnamurthi. FrTime: Functional Reactive Programming in PLT Scheme. Brown, CS{-}03{-}20, 2004. \href{https://cs.brown.edu/research/pubs/techreports/reports/CS-03-20.html}{\Snolinkurl{https://cs.brown.edu/research/pubs/techreports/reports/CS-03-20.html}}}}} \\
\hbox{\Autocolbibnumber{[7]}} &
\hbox{\Autobibtarget{\label{t:x28autobib_x22Gregory_Hx2e_Cooper_and_Shriram_KrishnamurthiEmbedding_Dynamic_Dataflow_in_a_Callx2dbyx2dValue_LanguageIn_Procx2e_15th_European_Conference_on_Programming_Languages_and_Systemsx2c_ESOPx2706x2c_ppx2e_294x2dx2d3082006doix3a10x2e1007x2f11693024x5f20x22x29}\Autocolbibentry{Gregory H. Cooper and Shriram Krishnamurthi. Embedding Dynamic Dataflow in a Call{-}by{-}Value Language. In \textit{Proc. 15th European Conference on Programming Languages and Systems}, ESOP{'}06, pp. 294{--}308, 2006. \pseudodoi{doi:\href{https://doi.org/10.1007/11693024_20}{10{\hbox{\texttt{.}}}1007/11693024{\char`\_}20}}}}} \\
\hbox{\Autocolbibnumber{[8]}} &
\hbox{\Autobibtarget{\label{t:x28autobib_x22Gregory_Harold_CooperIntegrating_Dataflow_Evaluation_into_a_Practical_Higherx2dOrder_Callx2dbyx2dValue_LanguagePhD_dissertationx2c_Brown_University2008httpsx3ax2fx2fcsx2ebrownx2eedux2fpeoplex2fghcooperx2fthesisx2epdfx22x29}\Autocolbibentry{Gregory Harold Cooper. Integrating Dataflow Evaluation into a Practical Higher{-}Order Call{-}by{-}Value Language. PhD dissertation, Brown University, 2008. \href{https://cs.brown.edu/people/ghcooper/thesis.pdf}{\Snolinkurl{https://cs.brown.edu/people/ghcooper/thesis.pdf}}}}} \\
\hbox{\Autocolbibnumber{[9]}} &
\hbox{\Autobibtarget{\label{t:x28autobib_x22Antony_CourtneyFrappxe9x3a_Functional_Reactive_Programming_in_JavaIn_Procx2e_Practical_Aspects_of_Declarative_Languages2001httpsx3ax2fx2fdoix2eorgx2f10x2e1007x2f3x2d540x2d45241x2d9x5f3x22x29}\Autocolbibentry{Antony Courtney. Frapp\'{e}: Functional Reactive Programming in Java. In \textit{Proc. Practical Aspects of Declarative Languages}, 2001. \href{https://doi.org/10.1007/3-540-45241-9_3}{\Snolinkurl{https://doi.org/10.1007/3-540-45241-9_3}}}}} \\
\hbox{\Autocolbibnumber{[10]}} &
\hbox{\Autobibtarget{\label{t:x28autobib_x22Evan_CzaplickiElm_x2d_delightful_language_for_reliable_web_applications2021httpsx3ax2fx2felmx2dlangx2eorgRetrieved_June_2023x2ex22x29}\Autocolbibentry{Evan Czaplicki. Elm {-} delightful language for reliable web applications. 2021. \href{https://elm-lang.org}{\Snolinkurl{https://elm-lang.org}} Retrieved June 2023.}}} \\
\hbox{\Autocolbibnumber{[11]}} &
\hbox{\Autobibtarget{\label{t:x28autobib_x22Robert_Bruce_Findlerx2c_John_Clementsx2c_Cormac_Flanaganx2c_Matthew_Flattx2c_Shriram_Krishnamurthix2c_Paul_Stecklerx2c_and_Matthias_FelleisenDrSchemex3a_A_programming_environment_for_Schemex2eJournal_of_Functional_Programming_12x282x29x2c_ppx2e_159x2dx2d1822002x22x29}\Autocolbibentry{Robert Bruce Findler, John Clements, Cormac Flanagan, Matthew Flatt, Shriram Krishnamurthi, Paul Steckler, and Matthias Felleisen. DrScheme: A programming environment for Scheme. \textit{Journal of Functional Programming} 12(2), pp. 159{--}182, 2002.}}} \\
\hbox{\Autocolbibnumber{[12]}} &
\hbox{\Autobibtarget{\label{t:x28autobib_x22Matthew_Flattx2c_Robert_Bruce_Findlerx2c_and_John_ClementsGUIx3a_Racket_Graphics_ToolkitPLT_Design_Incx2ex2c_PLTx2dTRx2d2010x2d32010httpsx3ax2fx2fracketx2dlangx2eorgx2ftr3x2fx22x29}\Autocolbibentry{Matthew Flatt, Robert Bruce Findler, and John Clements. GUI: Racket Graphics Toolkit. PLT Design Inc., PLT{-}TR{-}2010{-}3, 2010. \href{https://racket-lang.org/tr3/}{\Snolinkurl{https://racket-lang.org/tr3/}}}}} \\
\hbox{\Autocolbibnumber{[13]}} &
\hbox{\Autobibtarget{\label{t:x28autobib_x22Matthew_Flattx2c_Shriram_Krishnamurthix2c_and_Matthias_FelleisenClasses_and_mixinsIn_Procx2e_25th_ACM_SIGPLANx2dSIGACT_symposium_on_Principles_of_programming_languagesx2c_POPL_x2798x2c_ppx2e_171x2dx2d1831998doix3ahttpsx3ax2fx2fdoix2eorgx2f10x2e1145x2f268946x2e268961x22x29}\Autocolbibentry{Matthew Flatt, Shriram Krishnamurthi, and Matthias Felleisen. Classes and mixins. In \textit{Proc. 25th ACM SIGPLAN{-}SIGACT symposium on Principles of programming languages}, POPL {'}98, pp. 171{--}183, 1998. \pseudodoi{doi:\href{https://doi.org/https://doi.org/10.1145/268946.268961}{https{\hbox{\texttt{:}}}//doi{\hbox{\texttt{.}}}org/10{\hbox{\texttt{.}}}1145/268946{\hbox{\texttt{.}}}268961}}}}} \\
\hbox{\Autocolbibnumber{[14]}} &
\hbox{\Autobibtarget{\label{t:x28autobib_x22Matthew_Flatt_and_PLTReferencex3a_RacketPLT_Design_Incx2ex2c_PLTx2dTRx2d2010x2d12010httpsx3ax2fx2fracketx2dlangx2eorgx2ftr1x2fx22x29}\Autocolbibentry{Matthew Flatt and PLT. Reference: Racket. PLT Design Inc., PLT{-}TR{-}2010{-}1, 2010. \href{https://racket-lang.org/tr1/}{\Snolinkurl{https://racket-lang.org/tr1/}}}}} \\
\hbox{\Autocolbibnumber{[15]}} &
\hbox{\Autobibtarget{\label{t:x28autobib_x22Adele_GoldbergSmalltalkx2d80x3a_The_Interactive_Programming_EnvironmentAddisonx2dWesley_Publishersx2e1983x22x29}\Autocolbibentry{Adele Goldberg. \textit{Smalltalk{-}80: The Interactive Programming Environment}. Addison{-}Wesley Publishers., 1983.}}} \\
\hbox{\Autocolbibnumber{[16]}} &
\hbox{\Autobibtarget{\label{t:x28autobib_x22David_Goldbergx2c_Robert_Bruce_Findlerx2c_and_Matthew_FlattSuper_and_Innerx2dx2dx2dTogether_at_Lastx21In_Procx2e_Objectx2dOriented_Programmingx2c_Languagesx2c_Systemsx2c_and_Applications2004httpx3ax2fx2fwwwx2ecsx2eutahx2eedux2fpltx2fpublicationsx2foopsla04x2dgffx2epdfx22x29}\Autocolbibentry{David Goldberg, Robert Bruce Findler, and Matthew Flatt. Super and Inner{---}Together at Last! In \textit{Proc. Object{-}Oriented Programming, Languages, Systems, and Applications}, 2004. \href{http://www.cs.utah.edu/plt/publications/oopsla04-gff.pdf}{\Snolinkurl{http://www.cs.utah.edu/plt/publications/oopsla04-gff.pdf}}}}} \\
\hbox{\Autocolbibnumber{[17]}} &
\hbox{\Autobibtarget{\label{t:x28autobib_x22John_GruberDaring_Fireballx3a_Markdown2023httpsx3ax2fx2fdaringfireballx2enetx2fprojectsx2fmarkdownx2fRetrieved_June_2023x2ex22x29}\Autocolbibentry{John Gruber. Daring Fireball: Markdown. 2023. \href{https://daringfireball.net/projects/markdown/}{\Snolinkurl{https://daringfireball.net/projects/markdown/}} Retrieved June 2023.}}} \\
\hbox{\Autocolbibnumber{[18]}} &
\hbox{\Autobibtarget{\label{t:x28autobib_x22Daniel_Ignatoffx2c_Gregory_Hx2e_Cooperx2c_and_Shriram_KrishnamurthiCrossing_State_Linesx3a_Adapting_Objectx2dOriented_Frameworks_to_Functional_Reactive_LanguagesIn_Procx2e_Functional_and_Logic_Programmingx2c_FLOPS_20062006httpsx3ax2fx2flinkx2espringerx2ecomx2fchapterx2f10x2e1007x2f11737414x5f18x22x29}\Autocolbibentry{Daniel Ignatoff, Gregory H. Cooper, and Shriram Krishnamurthi. Crossing State Lines: Adapting Object{-}Oriented Frameworks to Functional Reactive Languages. In \textit{Proc. Functional and Logic Programming}, FLOPS 2006, 2006. \href{https://link.springer.com/chapter/10.1007/11737414_18}{\Snolinkurl{https://link.springer.com/chapter/10.1007/11737414_18}}}}} \\
\hbox{\Autocolbibnumber{[19]}} &
\hbox{\Autobibtarget{\label{t:x28autobib_x22Ralph_Ex2e_Johnson_and_Brian_FooteDesigning_Reusable_ClassesJournal_of_Objectx2dOriented_Programming_1x282x29x2c_ppx2e_22x2dx2d351988httpx3ax2fx2fwwwx2elaputanx2eorgx2fdrcx2fdrcx2ehtmlx22x29}\Autocolbibentry{Ralph E. Johnson and Brian Foote. Designing Reusable Classes. \textit{Journal of Object{-}Oriented Programming} 1(2), pp. 22{--}35, 1988. \href{http://www.laputan.org/drc/drc.html}{\Snolinkurl{http://www.laputan.org/drc/drc.html}}}}} \\
\hbox{\Autocolbibnumber{[20]}} &
\hbox{\Autobibtarget{\label{t:x28autobib_x22Dx2e_Ben_Knoblefrosthavenx2dmanager2022httpsx3ax2fx2fgithubx2ecomx2fbenknoblex2ffrosthavenx2dmanagerx22x29}\Autocolbibentry{D. Ben Knoble. frosthaven{-}manager. 2022. \href{https://github.com/benknoble/frosthaven-manager}{\Snolinkurl{https://github.com/benknoble/frosthaven-manager}}}}} \\
\hbox{\Autocolbibnumber{[21]}} &
\hbox{\Autobibtarget{\label{t:x28autobib_x22Glenn_Ex2e_Krasner_and_Stephen_Tx2e_PopeA_description_of_the_modelx2dviewx2dcontroller_user_interface_paradigm_in_the_Smalltalkx2d80_systemJournal_of_Objectx2dOriented_Programming_1x283x29x2c_ppx2e_26x2dx2d491988httpsx3ax2fx2fwwwx2eicsx2eucix2eedux2fx7eredmilesx2fics227x2dSQ04x2fpapersx2fKrasnerPope88x2epdfx22x29}\Autocolbibentry{Glenn E. Krasner and Stephen T. Pope. A description of the model{-}view{-}controller user interface paradigm in the Smalltalk{-}80 system. \textit{Journal of Object{-}Oriented Programming} 1(3), pp. 26{--}49, 1988. \href{https://www.ics.uci.edu/~redmiles/ics227-SQ04/papers/KrasnerPope88.pdf}{\Snolinkurl{https://www.ics.uci.edu/~redmiles/ics227-SQ04/papers/KrasnerPope88.pdf}}}}} \\
\hbox{\Autocolbibnumber{[22]}} &
\hbox{\Autobibtarget{\label{t:x28autobib_x22Leo_Ax2e_Meyerovichx2c_Arjun_Guhax2c_Jacob_Baskinx2c_Gregory_Hx2e_Cooperx2c_Michael_Greenbergx2c_Aleks_Bromfieldx2c_and_Shriram_KrishnamurthiFlapjaxx3a_a_programming_language_for_Ajax_applicationsIn_Procx2e_ACM_SIGPLAN_conference_on_Object_oriented_programming_systems_languages_and_applicationsx2c_OOPSLA_20092009httpsx3ax2fx2fdlx2eacmx2eorgx2fdoix2f10x2e1145x2f1640089x2e1640091x22x29}\Autocolbibentry{Leo A. Meyerovich, Arjun Guha, Jacob Baskin, Gregory H. Cooper, Michael Greenberg, Aleks Bromfield, and Shriram Krishnamurthi. Flapjax: a programming language for Ajax applications. In \textit{Proc. ACM SIGPLAN conference on Object oriented programming systems languages and applications}, OOPSLA 2009, 2009. \href{https://dl.acm.org/doi/10.1145/1640089.1640091}{\Snolinkurl{https://dl.acm.org/doi/10.1145/1640089.1640091}}}}} \\
\hbox{\Autocolbibnumber{[23]}} &
\hbox{\Autobibtarget{\label{t:x28autobib_x22David_Ax2e_MoonObjectx2doriented_programming_with_FlavorsIn_Procx2e_ACM_Conference_on_Objectx2doriented_Programmingx2c_Systemsx2c_Languagesx2c_and_Applicationsx2c_ppx2e_1x2dx2d81986httpsx3ax2fx2fwwwx2ecsx2etuftsx2eedux2fcompx2f150FPx2farchivex2fdavidx2dmoonx2fflavorsx2epdfx22x29}\Autocolbibentry{David A. Moon. Object{-}oriented programming with Flavors. In \textit{Proc. ACM Conference on Object{-}oriented Programming, Systems, Languages, and Applications}, pp. 1{--}8, 1986. \href{https://www.cs.tufts.edu/comp/150FP/archive/david-moon/flavors.pdf}{\Snolinkurl{https://www.cs.tufts.edu/comp/150FP/archive/david-moon/flavors.pdf}}}}} \\
\hbox{\Autocolbibnumber{[24]}} &
\hbox{\Autobibtarget{\label{t:x28autobib_x22Brad_Ax2e_MyersA_New_Model_for_Handling_InputACM_Transactions_on_Information_Systems_8x283x29x2c_ppx2e_289x2dx2d3201990doix3ahttpsx3ax2fx2fdoix2eorgx2f10x2e1145x2f98188x2e98204x22x29}\Autocolbibentry{Brad A. Myers. A New Model for Handling Input. \textit{ACM Transactions on Information Systems} 8(3), pp. 289{--}320, 1990. \pseudodoi{doi:\href{https://doi.org/https://doi.org/10.1145/98188.98204}{https{\hbox{\texttt{:}}}//doi{\hbox{\texttt{.}}}org/10{\hbox{\texttt{.}}}1145/98188{\hbox{\texttt{.}}}98204}}}}} \\
\hbox{\Autocolbibnumber{[25]}} &
\hbox{\Autobibtarget{\label{t:x28autobib_x22Andrew_Jx2e_Palayx2c_Wilfred_Jx2e_Hansenx2c_Michael_Lx2e_Kazarx2c_Mark_Shermanx2c_Maria_Gx2e_Wadlowx2c_Thomas_Px2e_Neuendorfferx2c_Zalman_Sternx2c_Miles_Baderx2c_and_Thom_PetersThe_Andrew_Toolkitx2014An_OverviewIn_Procx2e_USENIX_Winter_Conferencex2c_ppx2e_9x2dx2d221988x22x29}\Autocolbibentry{Andrew J. Palay, Wilfred J. Hansen, Michael L. Kazar, Mark Sherman, Maria G. Wadlow, Thomas P. Neuendorffer, Zalman Stern, Miles Bader, and Thom Peters. The Andrew Toolkit{---}An Overview. In \textit{Proc. USENIX Winter Conference}, pp. 9{--}22, 1988.}}} \\
\hbox{\Autocolbibnumber{[26]}} &
\hbox{\Autobibtarget{\label{t:x28autobib_x22Bogdan_PopaAnnouncing_GUI_Easy2021httpsx3ax2fx2fdefnx2eiox2f2021x2f08x2f01x2fannx2dguix2deasyx2fx22x29}\Autocolbibentry{Bogdan Popa. Announcing GUI Easy. 2021. \href{https://defn.io/2021/08/01/ann-gui-easy/}{\Snolinkurl{https://defn.io/2021/08/01/ann-gui-easy/}}}}} \\
\hbox{\Autocolbibnumber{[27]}} &
\hbox{\Autobibtarget{\label{t:x28autobib_x22racox3a_Racket_Commandx2dLine_Tools2010httpsx3ax2fx2fdocsx2eracketx2dlangx2eorgx2fracox2findexx2ehtmlx22x29}\Autocolbibentry{raco: Racket Command{-}Line Tools. 2010. \href{https://docs.racket-lang.org/raco/index.html}{\Snolinkurl{https://docs.racket-lang.org/raco/index.html}}}}} \\
\hbox{\Autocolbibnumber{[28]}} &
\hbox{\Autobibtarget{\label{t:x28autobib_x22reagentx2dprojectReagent2023httpsx3ax2fx2fgithubx2ecomx2freagentx2dprojectx2freagentRetrieved_June_2023x2ex22x29}\Autocolbibentry{reagent{-}project. Reagent. 2023. \href{https://github.com/reagent-project/reagent}{\Snolinkurl{https://github.com/reagent-project/reagent}} Retrieved June 2023.}}} \\
\hbox{\Autocolbibnumber{[29]}} &
\hbox{\Autobibtarget{\label{t:x28autobib_x22Meta_Open_SourceReact2023httpsx3ax2fx2freactx2edevRetrieved_June_2023x2ex22x29}\Autocolbibentry{Meta Open Source. React. 2023. \href{https://react.dev}{\Snolinkurl{https://react.dev}} Retrieved June 2023.}}} \\
\hbox{\Autocolbibnumber{[30]}} &
\hbox{\Autobibtarget{\label{t:x28autobib_x22Richard_Ex2e_SweetThe_Mesa_Programming_EnvironmentACM_SIGPLAN_Notices_20x287x29x2c_ppx2e_216x2dx2d2291985doix3ahttpsx3ax2fx2fdoix2eorgx2f10x2e1145x2f17919x2e806843x22x29}\Autocolbibentry{Richard E. Sweet. The Mesa Programming Environment. \textit{ACM SIGPLAN Notices} 20(7), pp. 216{--}229, 1985. \pseudodoi{doi:\href{https://doi.org/https://doi.org/10.1145/17919.806843}{https{\hbox{\texttt{:}}}//doi{\hbox{\texttt{.}}}org/10{\hbox{\texttt{.}}}1145/17919{\hbox{\texttt{.}}}806843}}}}} \\
\hbox{\Autocolbibnumber{[31]}} &
\hbox{\Autobibtarget{\label{t:x28autobib_x22Day_8_Technologyrex2dframe2023httpsx3ax2fx2fgithubx2ecomx2fday8x2frex2dframeRetrieved_June_2023x2ex22x29}\Autocolbibentry{Day 8 Technology. re{-}frame. 2023. \href{https://github.com/day8/re-frame}{\Snolinkurl{https://github.com/day8/re-frame}} Retrieved June 2023.}}} \\
\hbox{\Autocolbibnumber{[32]}} &
\hbox{\Autobibtarget{\label{t:x28autobib_x22Donald_Cx2e_WallaceTajo_Functional_Specification_Version_6x2e0x2cXerox_Internal_Document1980x22x29}\Autocolbibentry{Donald C. Wallace. Tajo Functional Specification Version 6.0, Xerox Internal Document, 1980.}}}\end{bigtabular}

\postDoc
\end{document}